\documentclass[journal]{IEEEtran}
\usepackage{xcolor,soul,framed} 
\usepackage{cite}
\usepackage{amsmath,amssymb,amsfonts}
\usepackage{graphicx}
\usepackage{textcomp}
\usepackage{dcolumn}
\usepackage{bm}
\usepackage{subfigure} 
\usepackage{pict2e}
\usepackage{epstopdf}
\usepackage{epstopdf}
\usepackage{amsmath}
\usepackage{arydshln}
\usepackage[hyphens]{url}

\ifCLASSINFOpdf
\else
\fi

\begin{document}

\onecolumn
\newpage
\thispagestyle{empty}

\textbf{Copyright:}\\
\copyright 2020 IEEE. Personal use of this material is permitted.  Permission from IEEE must be obtained for all other uses, in any current or future media, including reprinting/republishing this material for advertising or promotional purposes, creating new collective works, for resale or redistribution to servers or lists, or reuse of any copyrighted component of this work in other works.\\

\textbf{Disclaimer:}\\
This work has been published on \textit{IEEE Transactions on Microwave Theory and Techniques}. DOI: 10.1109/TMTT.2020.3021087\\

\newpage

\title{Dispersion and Filtering Properties of Rectangular Waveguides Loaded with Holey Structures}

\author{\'{A}ngel Palomares-Caballero, Antonio Alex-Amor, Pablo Padilla, and Juan F. Valenzuela-Vald\'{e}s
\thanks{This  work  was   sup-ported  in  part  by  the  Spanish  Research  and  Development  National  Program under Project TIN2016-75097-P, Project RTI2018-102002-A-I00, and Project EQC2018-004988-P;  in  part  by  the  “Junta  de  Andalucía”  under  Project  B-TIC-402-UGR18  and  Project  P18.RT.4830;  in  part  by  the  Predoctoral  Grant FPU18/01965;  in  part  by  the  CEIMar  Project  CEIJ-020;  and  in  part  by  the University  of  Granada  for  Young  Researchers  under  Project  PPJIA2019.10. \textit{(Corresponding author: \'{A}ngel Palomares-Caballero.)}}
\thanks{\'{A}. Palomares-Caballero, A. Alex-Amor, Juan Valenzuela-Vald\'{e}s and P. Padilla are with the Departamento de Teor\'{i}a de la Se\~{n}al, Telem\'{a}tica y Comunicaciones, Universidad de Granada, 18071 Granada, Spain (e-mail: angelpc@ugr.es; aalex@ugr.es; juanvalenzuela@ugr.es; pablopadilla@ugr.es).}
\thanks{A. Alex-Amor is with the Information Processing and Telecommunications Center, Universidad Polit\'{e}cnica de Madrid, 28040 Madrid, Spain (e-mail: aalex@gr.ssr.upm.es).}
}

\twocolumn

\maketitle
\newcommand*{\bigs}[1]{\vcenter{\hbox{\scalebox{2}[8.2]{\ensuremath#1}}}}

\newcommand*{\bigstwo}[1]{\vcenter{\hbox{ \scalebox{1}[4.4]{\ensuremath#1}}}}

\begin{abstract}
This paper analyzes thoroughly the dispersion and filtering features of periodic holey waveguides in the millimeter-wave frequency range. Two structures are mainly studied depending on the glide and mirror symmetries of the holes. A parametric study of the dispersion characteristics of their unit cells is carried out. Glide-symmetric holey waveguides provide a higher propagation constant and a low dispersion over a wide frequency range regarding hollow waveguides. This property is particularly useful for the design of low-loss and low-dispersive phase shifters. We also demonstrate that glide-symmetric holey waveguides are less dispersive than waveguides loaded with glide-symmetric pins. Furthermore, we perform a Bloch analysis to compute the attenuation constants in holey waveguides with mirror and broken glide symmetries. Both configurations are demonstrated to be suitable for filter design. Finally, the simulation results are validated with two prototypes in gap-waveguide technology. The first one is a 180\textsuperscript{o} phase shifter based on a glide-symmetric holey configuration that achieves a flat phase shift response over a wide frequency range (27.5\% frequency bandwidth). The second one is a filter based on a mirror-symmetric holey structure with 20-dB rejection from 63 GHz to 75 GHz. 
\end{abstract}

\vspace{6pt}

\begin{IEEEkeywords}
Filter, gap waveguide, glide symmetry, hollow waveguide, low dispersion, millimeter-wave, mirror symmetry, multi-modal (M-M) transfer matrix, periodic holey structures, phase shifter.
\end{IEEEkeywords}

\IEEEpeerreviewmaketitle

\section{Introduction}

\IEEEPARstart{W}{aveguide} technology is one of the preferred guiding structures technology for high-frequency applications due to the low losses produced in the field propagation \cite{Pozar}. Hollow waveguides are suitable for the design of electronic devices that operate at millimeter-wave and terahertz regimes \cite{WaveguideTHz}, at the expense of the complexities in its manufacturing process that were lately relieved with the implementation of gap waveguides \cite{GapWaveguide1, GapWaveguide2,GapWaveguide3, GapWaveguide4, GapWaveguide5}.

In communication systems, the implementation of filters and phase shifters is typically necessary \cite{MultibeamAntennas,CommSystem1,CommSystem2}. A strategy to design these RF (radio-frequency) components is the inclusion of periodic structures inside the waveguide \cite{LoadedWaveguide_theory1,LoadedWaveguide_theory2,LoadedWaveguide_theory3}. In the case of filters, the stopbands produced by the periodic configuration are utilized to tune rejection bands \cite{LoadedWaveguide_filter1,LoadedWaveguide_filter2,LoadedWaveguide_filter3,LoadedWaveguide_filter4,LoadedWaveguide_filter5,LoadedWaveguide_filter6, LoadedWaveguide_filter7, LoadedWaveguide_filter8}. The design of phase shifters based on metallo-dielectric periodic structures allows to artificially modify the propagation constant in the waveguide with high tunability. In substrate integrate waveguide (SIW) technology, there are some designs that make use of periodic metallic posts \cite{SIW_phaseShifter1}, omega particles \cite{SIW_phaseShifter2} or thin slots \cite{SIW_phaseShifter3} to achieve the desired phase shift. Conversely, only a few phase shifters based on periodic structures are reported in hollow waveguides. Corrugations \cite{WaveguidePhaseShifter_1,WaveguidePhaseShifter_2} and structures based on pins \cite{PinPhaseShifter_1,PinPhaseShifter_2,Hedgehog_phaseShifter} are commonly used.

Some previous works have already combined periodic structures with glide symmetry to enhance the properties of their devices: increase the propagation constant and reduce the dispersion over a wide frequency range \cite{App_HighSymmetry0,App_HighSymmetry1,App_HighSymmetry2,App_HighSymmetry3,App_HighSymmetry4,App_HighSymmetry5, App_HighSymmetry6, App_HighSymmetry7, Rev3_1, Rev3_2}. Glide-symmetric holey structures \cite{Rev1} are of particular interest from a manufacturing point of view due to their cost-effective fabrication with milling techniques. For instance, glide-symmetric holes have been implemented in a parallel-plate waveguide \cite{GlideSymm_holes1} and in multilayer waveguide \cite{GlideSymm_holes1_2}. A similar periodic structure with a higher separation between plates is used in \cite{GlideSymm_holes2} to design a prism in order to decrease the beam squint of a leaky-wave antenna. Furthermore, it has been detailed in \cite{GlideSymm_holes3} that breaking the glide symmetry is a complementary tool for the design of filters, due to the creation of a stopband. The only reported work that uses glide-symmetric holey structures, namely braided glide symmetry, for the design of a filter device in waveguide is \cite{GlideSymm_holes4}.

Most of the previously reported works are focused on the study of parallel-plate structures loaded with glide-symmetric holes. In this paper, we analyze the wave propagation when periodic holes possessing glide and mirror symmetries are inserted in the upper and lower plates of the rectangular waveguide. We characterize the influence of the geometrical parameters that define the holey structures on their dispersion curves. In particular, the variations of the cutoff frequencies in the holey waveguide are studied in detail. A comparison between waveguides loaded with glide-symmetric holes and pins \cite{PinPhaseShifter_2} is also carried out, showing that holey waveguides are generally less dispersive than pin-loaded waveguides. In addition, the attenuation constant is computed to accurately characterize the stopbands that appear in both holey configurations. Finally, two prototypes are manufactured for validation purposes: a wideband and low-dispersive phase shifter based on a glide-symmetric holey configuration that operates in V-band, and a filter based on the a mirror-symmetric holey configuration with 20-dB rejection from 63 GHz to 75 GHz.

The paper is organized as follows. In Section II, the analysis of the dispersion properties of the holey waveguide is carried out. We also elaborate a comparison with pin-loaded waveguides, as well as a discussion of the stopband properties for the mirror-symmetric and broken glide-symmetric cases. Section III presents the design, manufacturing and experimental results of a filter and a wideband phase shifter composed by the analyzed mirror-symmetric and glide-symmetric holey unit cells, respectively. Finally, the conclusions are drawn in Section IV.

\section{Dispersion and stopband analysis of the holey waveguide}

\begin{figure}[t!]
	\centering
	\subfigure[]{\includegraphics[width= 0.49\textwidth]{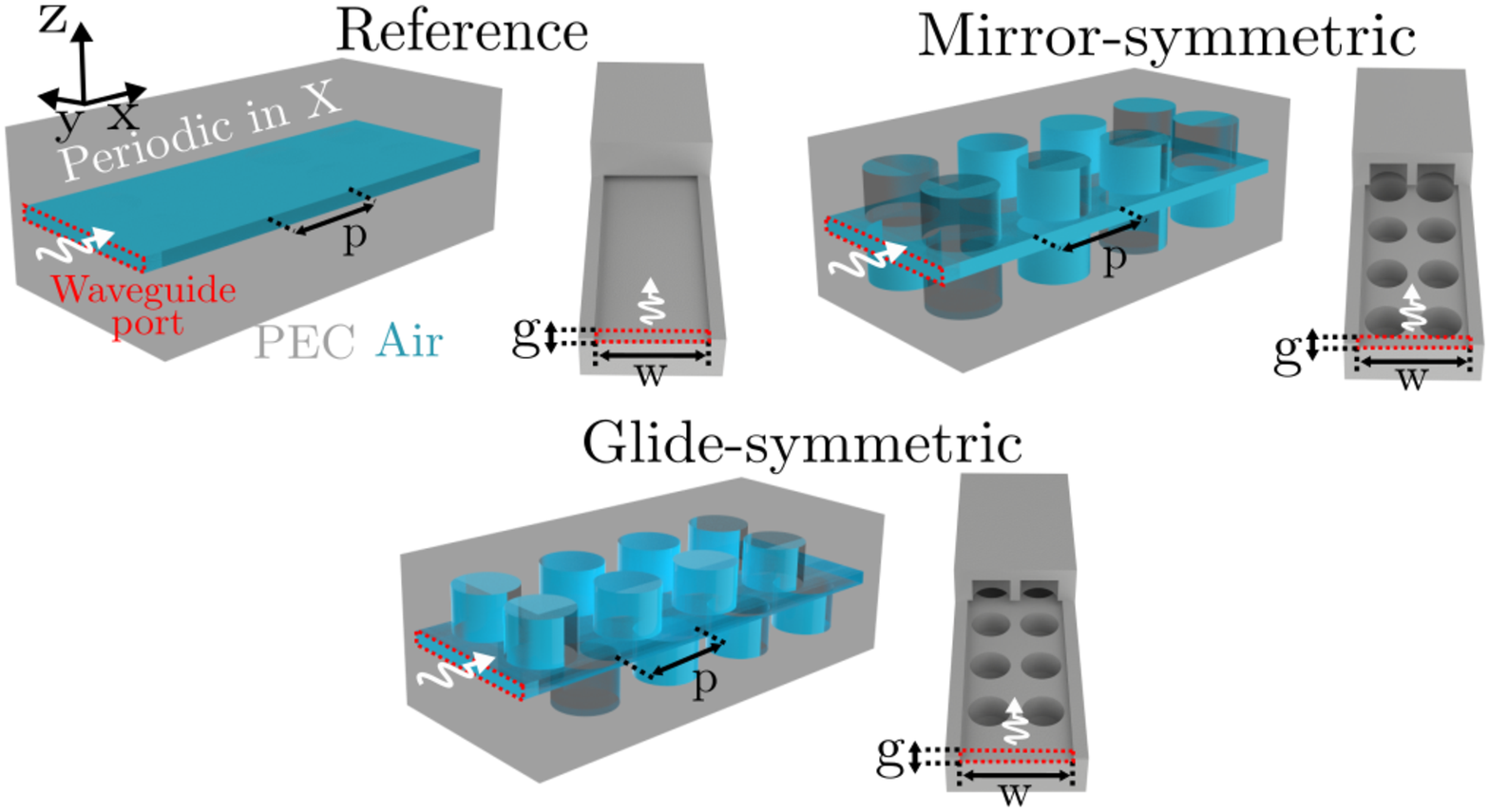}\label{figure1a}}
	\subfigure[]{\includegraphics[width= 0.48\textwidth]{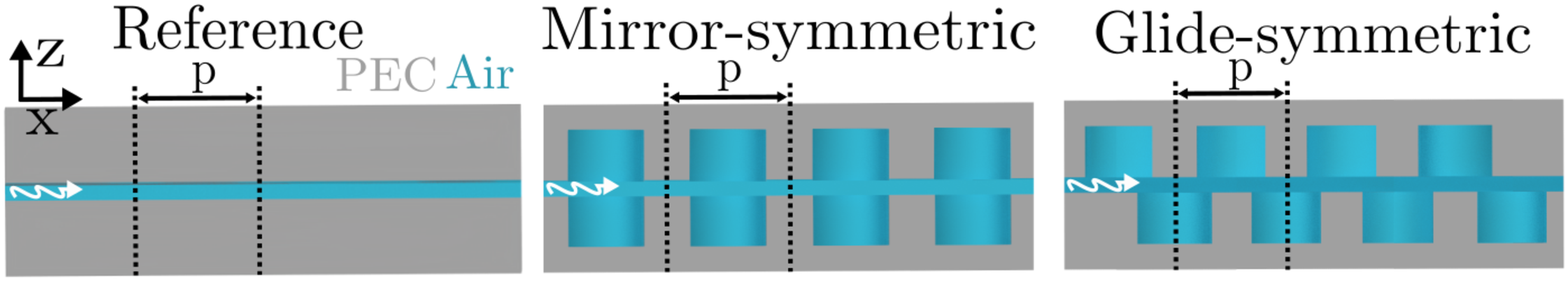}\label{figure1b}}
	\subfigure[]{\includegraphics[width= 0.48\textwidth]{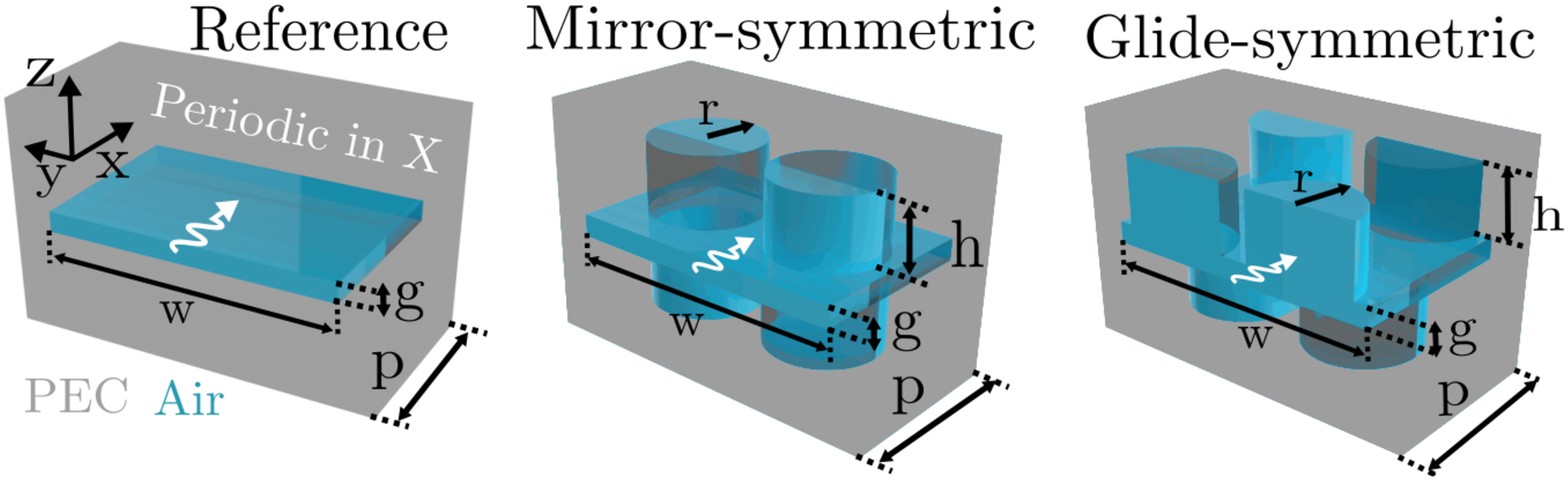}\label{figure1c}}
	\hspace{1mm}
	\caption{Reference and holey waveguide periodic structures. (a) Formed by four unit cells where input waveguide is marked in red. (b) Longitudinal cut views and (c) Unit cells.} 
	\label{figure1}
\end{figure}

In this section, we study the wave propagation in periodic holey waveguides and analyze their dispersion and stopband features. The periodic holey waveguides under study are illustrated in Fig. \ref{figure1a} where perfect metallic boundary conditions are enforced on the lateral walls of the waveguide. In this figure, the air zones inside the periodic structures are depicted to observe the configurations of the holes inside the waveguide. A cut showing their longitudinal sections is also presented in Fig. \ref{figure1b}. Two configurations are considered depending on the symmetry of the holes of the upper and lower waveguide plates. In the mirror-symmetric configuration, the upper and lower holes are aligned along the \textit{z}-axis. In the glide-symmetric configuration, the lower holes are off-shifted half a period in \textit{x}-direction in regard to the upper holes \cite{App_HighSymmetry1}. A reference waveguide is also considered in the study in order to compare the properties of the holey unit cells to it.

Fig. \ref{figure1c} illustrates the unit cells associated to the aforementioned periodic holey 1D structures. Both mirror-symmetric and glide-symmetric structures are periodic in \textit{x}-direction. They have a waveguide height \textit{g} and a waveguide width \textit{w} whose size corresponds to the WR15 waveguide. All simulated results in this section are related to lossless structures. The computation of the dispersion diagrams is performed in commercial software \textit{CST Microwave Studio} and an own code based on multi-modal analysis. Our reference unit cell has the following dimensions: height of the waveguide $g = 0.3$ mm, width of the waveguide $w = 3.76$ mm and period $\textit{p} = 2.4$ mm. Moreover, for the sake of completeness, the first or fundamental and second propagating modes of the unit cell are represented in the dispersion diagrams.

\subsection{Dispersion Properties}

In this subsection, the effect of the geometrical parameters in the holey unit cells is analyzed in detail and compared to the reference unit cell. Both mirror-symmetric and glide-symmetric unit cells are studied in order to identify their dispersion behavior and differences between them.  The present results are computed with the \textit{Eigensolver} of \textit{CST Microwave Studio}.

\begin{figure}[t!]
	\centering
	
	\subfigure[]{\includegraphics[width= 0.234\textwidth]{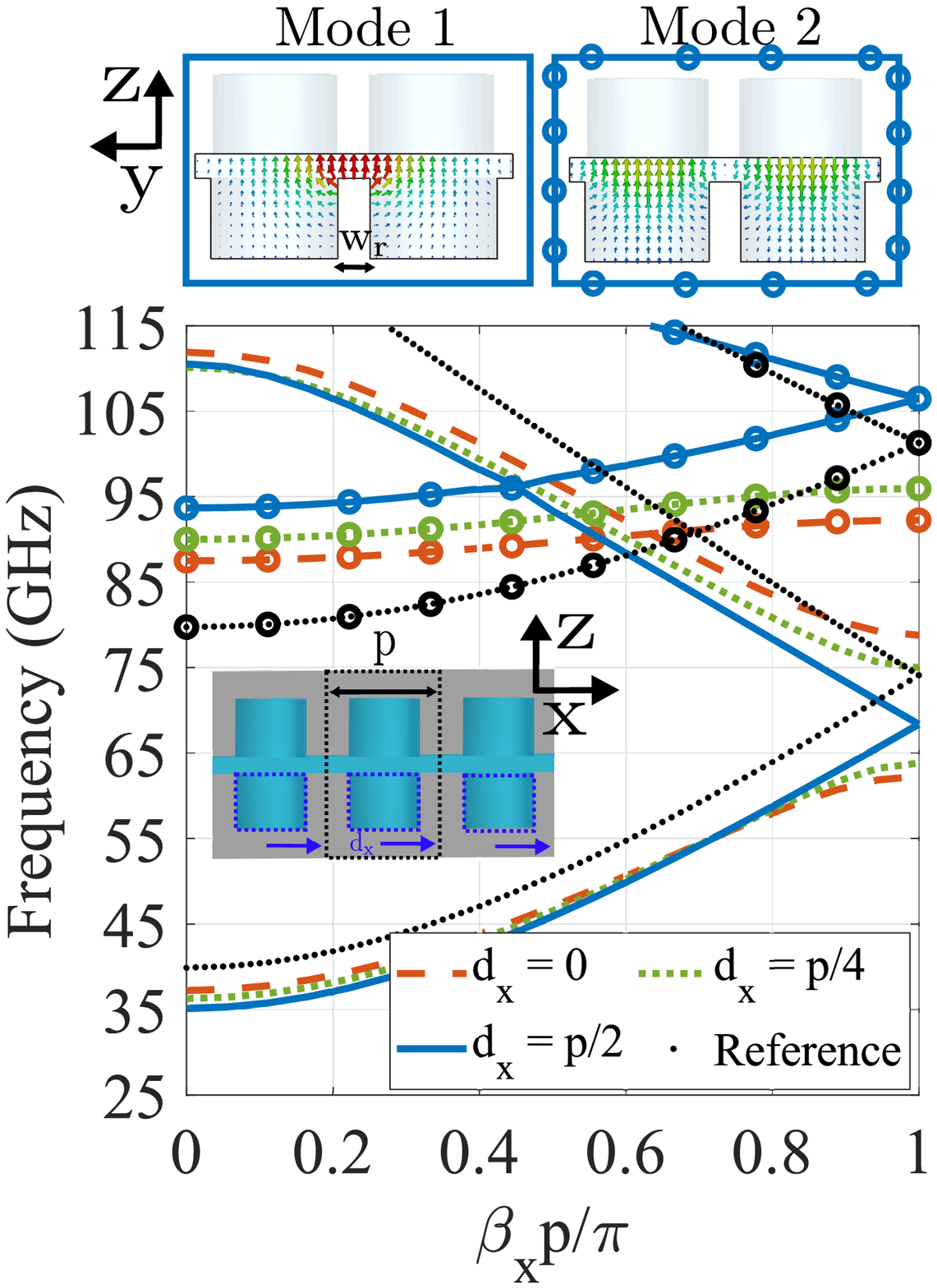}\label{figure2a}}
	\hspace{0.5mm}
	\subfigure[]{\includegraphics[width= 0.239\textwidth]{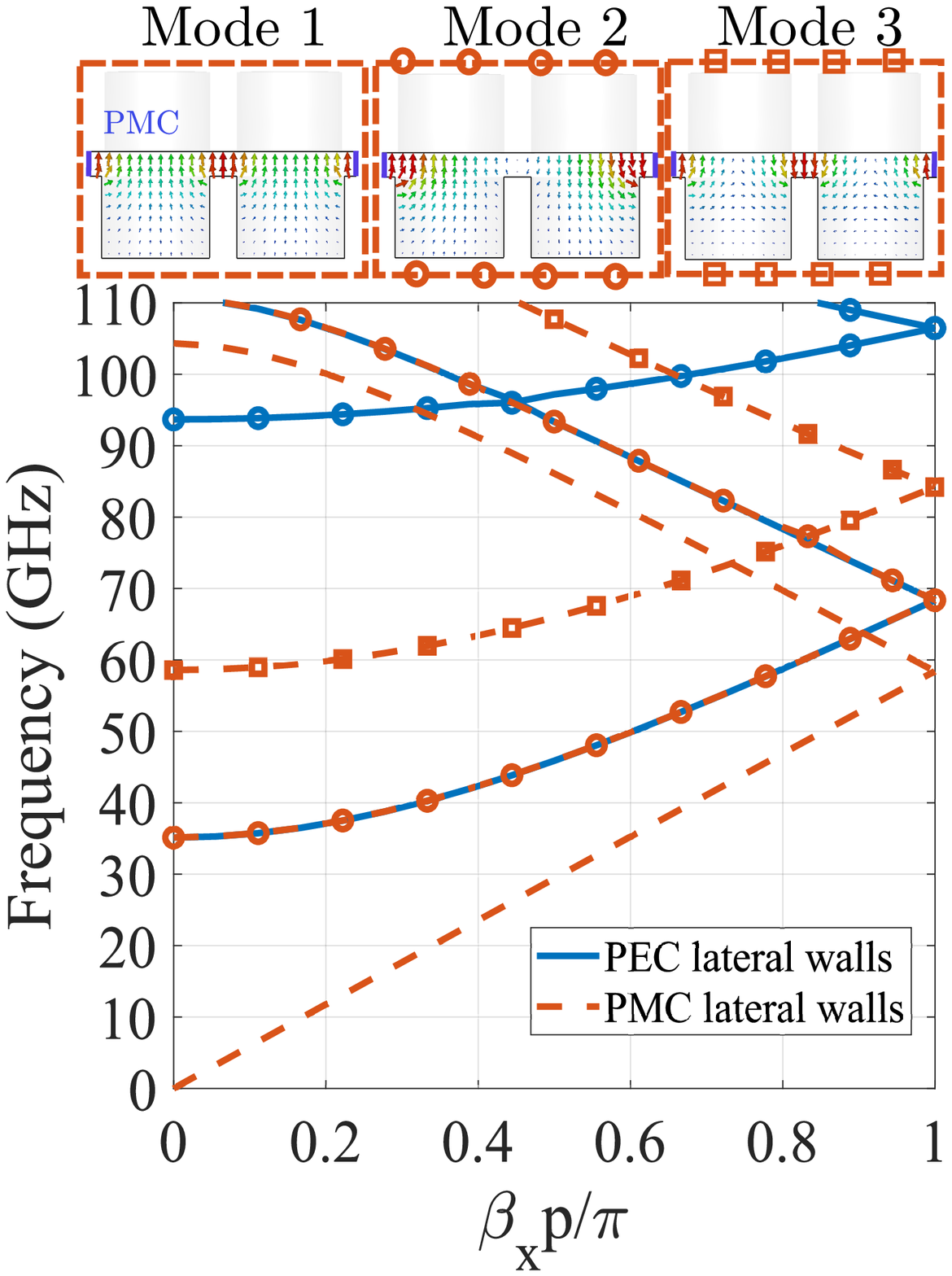}\label{figure2b}}
	
	\caption{(a) Dispersion diagram of a glide-symmetric unit cell when the lower holes are displaced \textit{d\textsubscript{x}} in \textit{x}-direction. (b) Dispersion diagram when PEC or PMC are placed at lateral walls in the glide-symmetric unit cell. First, second and third propagating modes are displayed in the figure. Holey waveguide unit cell has the following dimensions: $p = 2.4$ mm, $g = 0.3$ mm, $h = 1$ mm and $r = 0.75$ mm.}
	\label{figure2}
\end{figure}

Fig. \ref{figure2a} presents the dispersion diagram of a holey waveguide whose lower holes are off-shifted a distance $d_x$ with respect to the upper holes. Displacements of $d_x=0$ and $d_x=p/2$ correspond to the mirror-symmetric and glide-symmetric configuration, respectively. A displacement different from the glide-symmetric one introduces stopbands. This effect is also appreciated for other kinds of unit cells \cite{Rev3_1,App_HighSymmetry4}. Another advantage provided by the glide-symmetric configuration is that the first mode is approximately parallel regarding the reference waveguide in a wide frequency range. This dispersive behavior will be used in Section III to design a low-dispersive phase shifter in waveguide. For the sake of completeness, Fig. \ref{figure2b} presents the comparison of the glide-symmetric unit cell with PEC (perfect electric conductor) and PMC (perfect magnetic conductor) lateral walls. In the case of placing a lateral PMC condition, a mode without cutoff frequency propagates. This mode corresponds to the TEM mode in a parallel-plate waveguide.\\

\begin{figure}[t!]
	\centering
	
	\subfigure[]{\includegraphics[width= 0.415\textwidth]{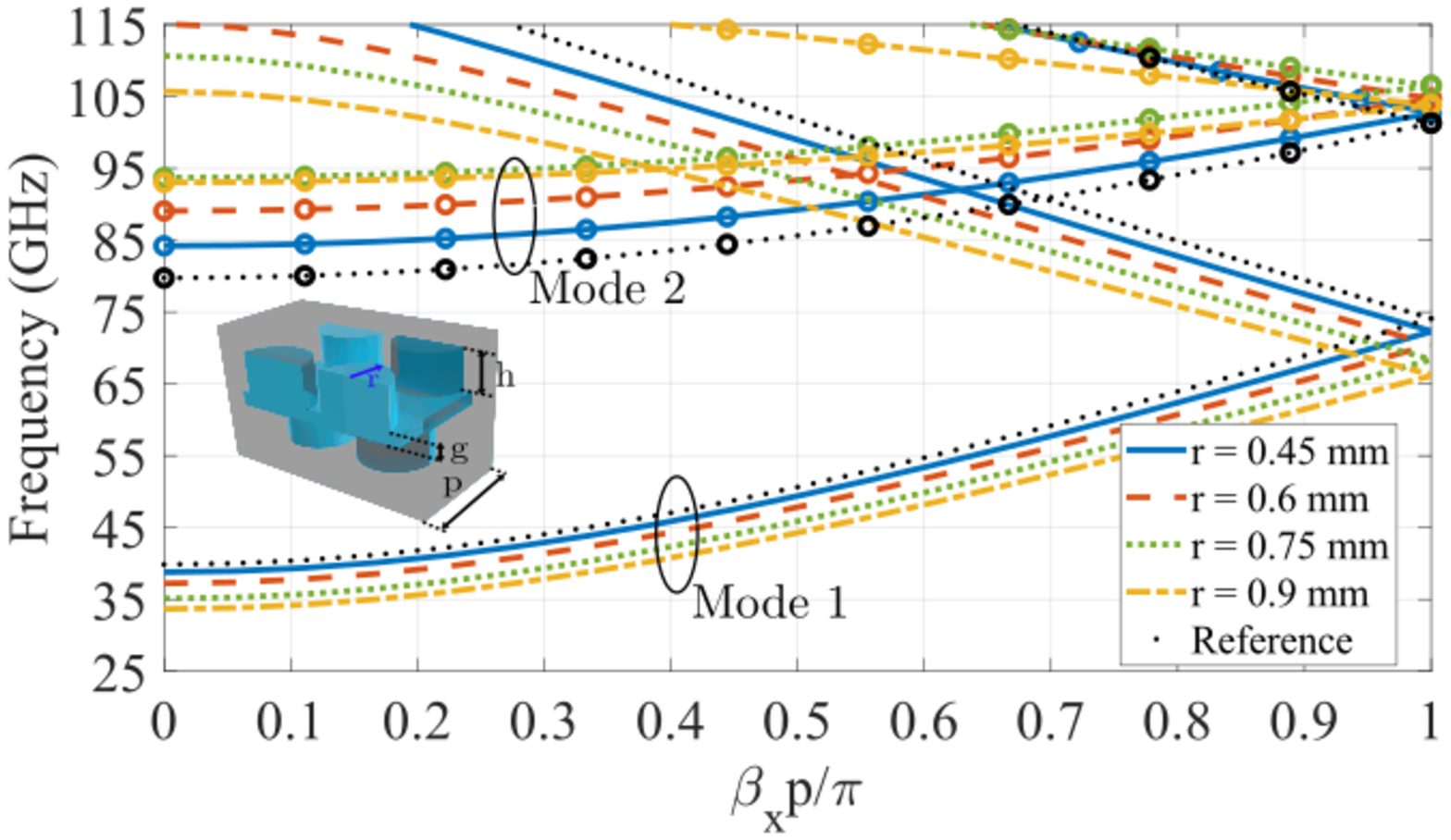}\label{figure3a}}
	\hspace{1mm}
	\subfigure[]{\includegraphics[width= 0.415\textwidth]{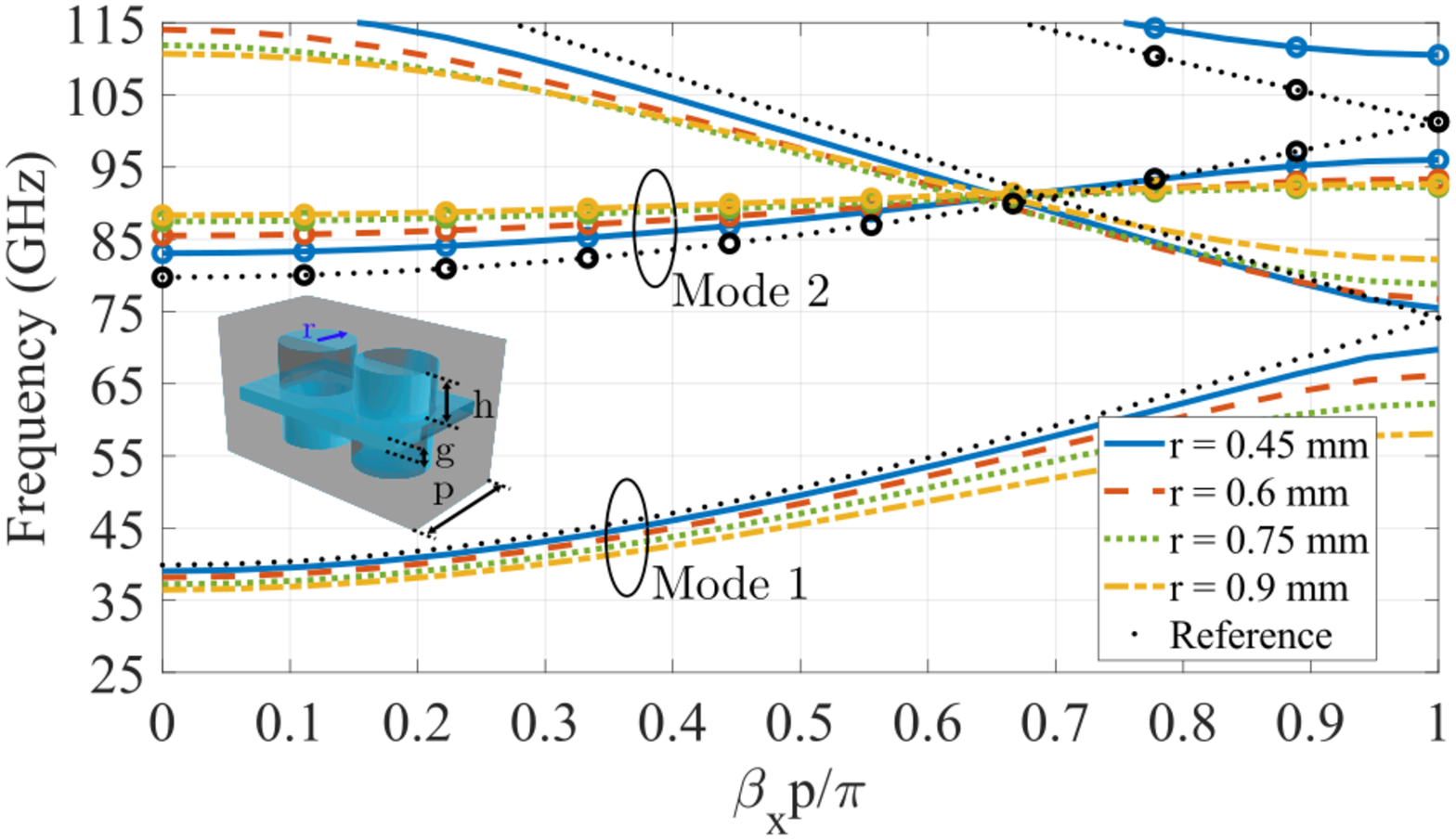}\label{figure3b}}
	
	\caption{Dispersion diagrams of the (a) Glide-symmetric and (b) Mirror-symmetric holey waveguides when the radius of the holes is modified. First and second propagating modes are displayed in the figure. The geometrical parameters of the depicted unit cells are: $p = 2.4$ mm, $g = 0.3$ mm, $h = 1$ mm and $w = 3.76$ mm.}
	\label{figure3}
\end{figure}

The influence of hole radius \textit{r} on the dispersion diagram of the glide-symmetric and mirror-symmetric unit cells is illustrated in Fig. \ref{figure3}. As previously evidenced in Fig. \ref{figure2}, a stopband exists between the first and second propagating modes when the holes do not preserve glide symmetry. In both structures, an increase in the radius of the hole provokes a denser unit cell, in terms of propagation constant, due to a higher interaction between the propagating modes and the hole radius. Another important effect that is observed is the increase in the separation between the cutoff frequencies of the first and second propagating modes. 

\begin{figure}[t!]
	\centering
	\subfigure[]{\includegraphics[width= 0.415\textwidth]{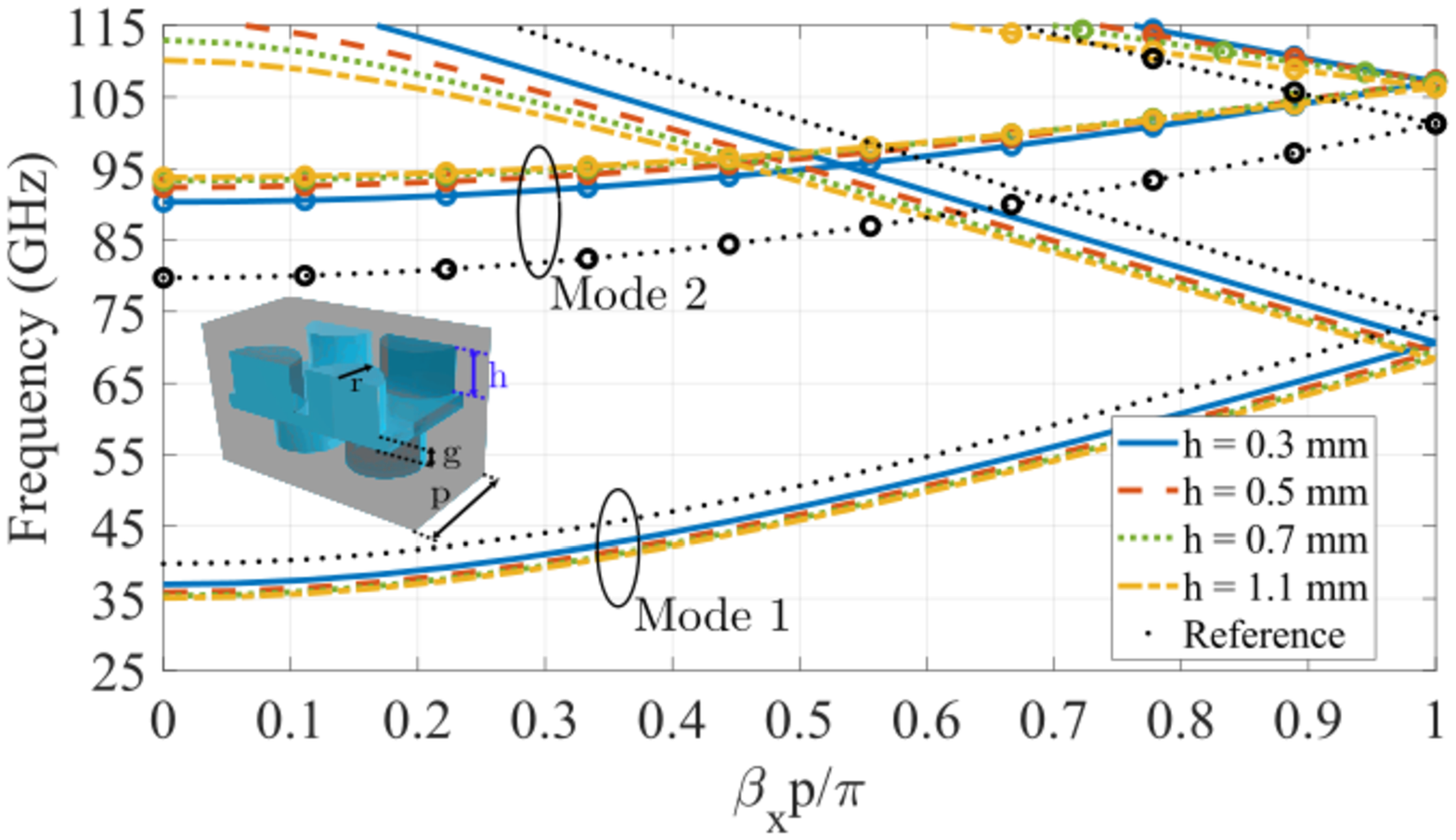}\label{figure4a}}
	\hspace{1mm}
	\subfigure[]{\includegraphics[width= 0.415\textwidth]{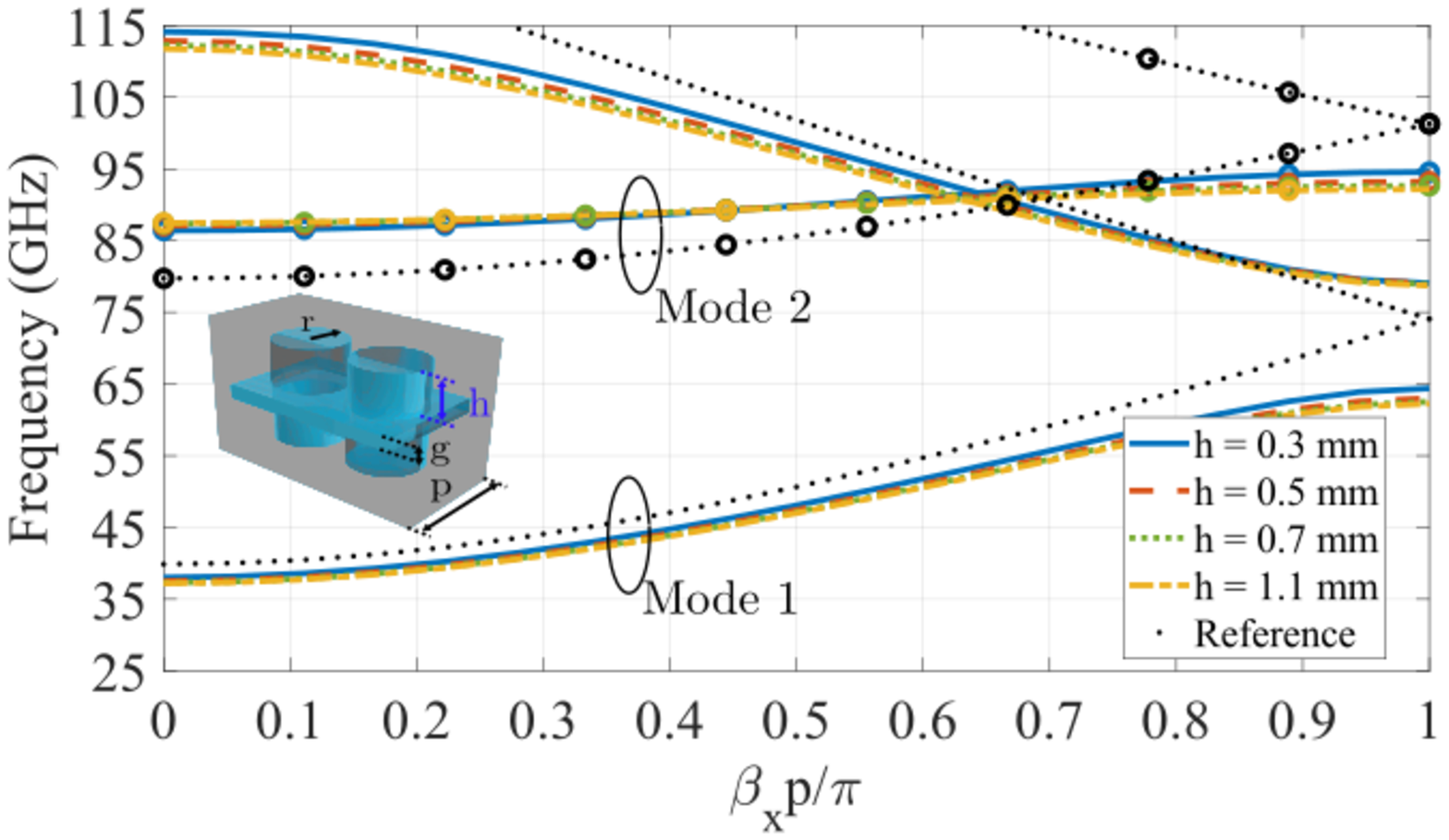}\label{figure4b}}
	
	\caption{Dispersion diagrams of the (a) Glide-symmetric and (b) Mirror-symmetric holey waveguides when varying the depth of the holes. First and second propagating modes are displayed in the figure. The geometrical parameters of the unit cells are: \textit{p} = 2.4 mm, \textit{g} = 0.3 mm and \textit{r} = 0.75 mm.}
	\label{figure4}
\end{figure}

Fig. \ref{figure4} illustrates the dispersion diagram for the glide-symmetric and mirror-symmetric structures when the depth of the hole \textit{h} is modified. In both configurations, an increase in the depth of the hole provokes a denser unit cell. Nevertheless, from certain hole depth, the dispersion diagram remains unchanged since the modes are evanescent inside the holes. This effect also occurs in holey metasurfaces \cite{App_HighSymmetry2,App_HighSymmetry5,App_HighSymmetry7,Rev1}. This enables an interesting cost-effective manufacturing of waveguide structures with holey unit cells by through holes if the selected hole depth is large enough. That is, the end of the hole does not need to be short-circuited if the hole is depth in terms of wavelength.

\begin{figure}[t!]
	\centering
	\subfigure[]{\includegraphics[width= 0.415\textwidth]{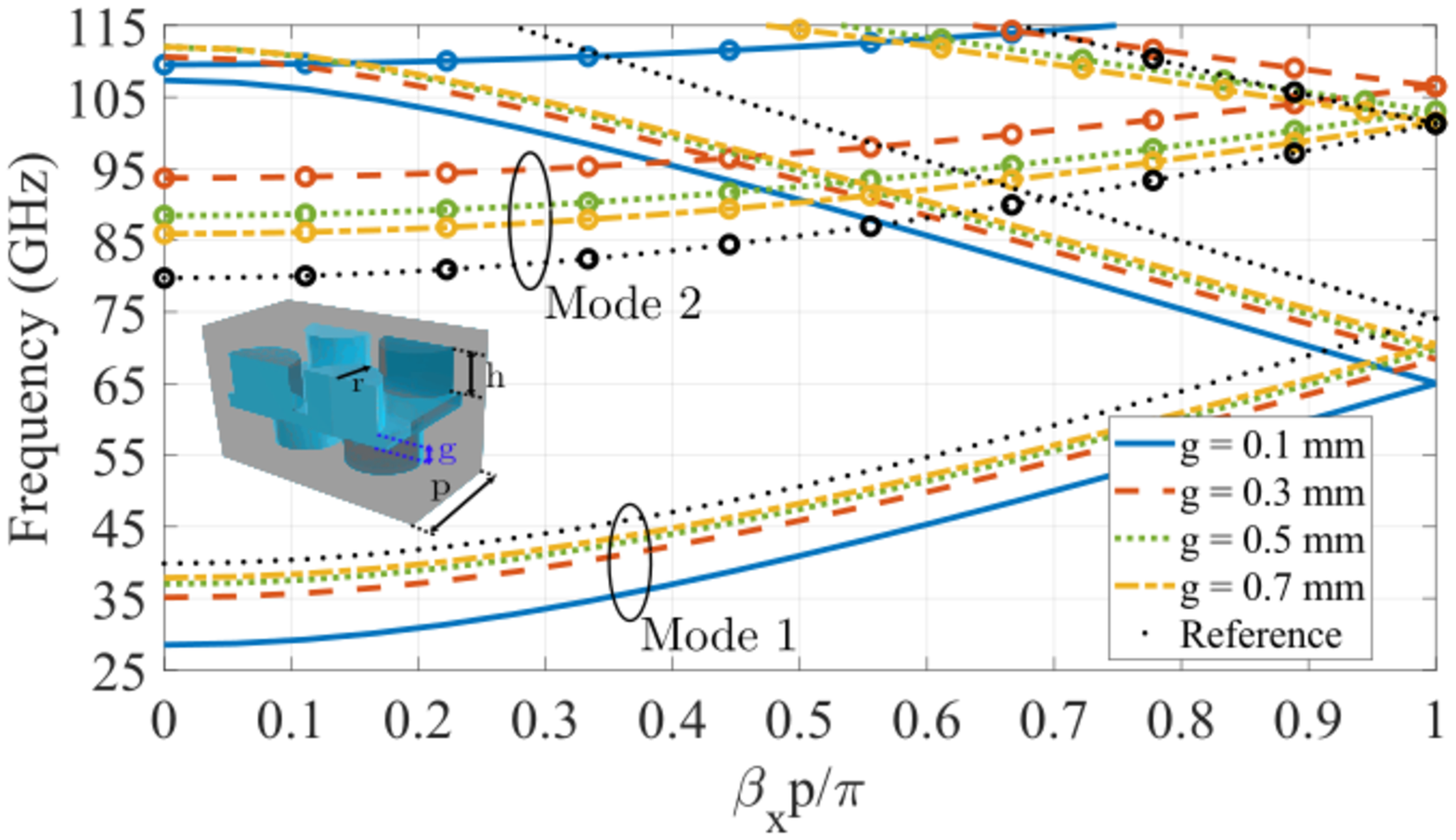}\label{figure5a}}
	\hspace{1mm}
	\subfigure[]{\includegraphics[width= 0.415\textwidth]{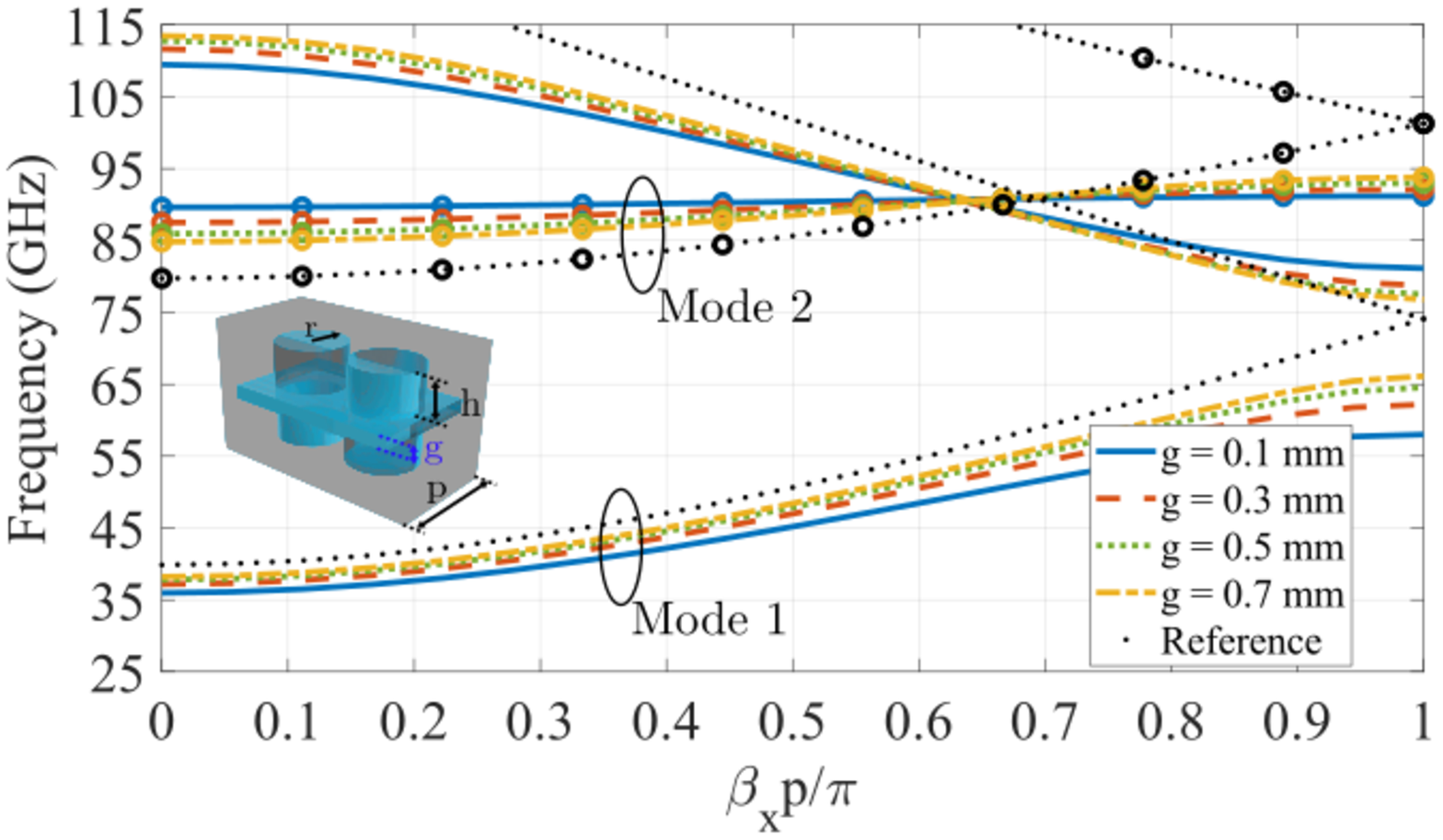}\label{figure5b}}
	
	\caption{Dispersion diagrams of the (a) Glide-symmetric and (b) Mirror-symmetric holey waveguides when varying the height of the waveguide. First and second propagating modes are displayed in the figure. The geometrical parameters of the unit cells are: \textit{p} = 2.4 mm, \textit{r} = 0.75 mm and \textit{h} = 1 mm.}
	\label{figure5}
\end{figure}

In Fig. \ref{figure5}, the influence of the height of the waveguide \textit{g} is analyzed. It is observed that smaller values of \textit{g} provide a higher propagation constant, since the top and bottom holes are closer and their interaction with the propagating mode is stronger. This fact makes the need of a waveguide height small enough in order to take into account the effect of the holes. An additional effect of reducing the height of the waveguide is the increase in the separation of the cutoff frequencies, similar to the one shown in Fig. \ref{figure3}. This is another possibility to extend the bandwidth of the fundamental mode.

\begin{figure}[t]
	\centering
	\subfigure[]{\includegraphics[width= 0.415\textwidth]{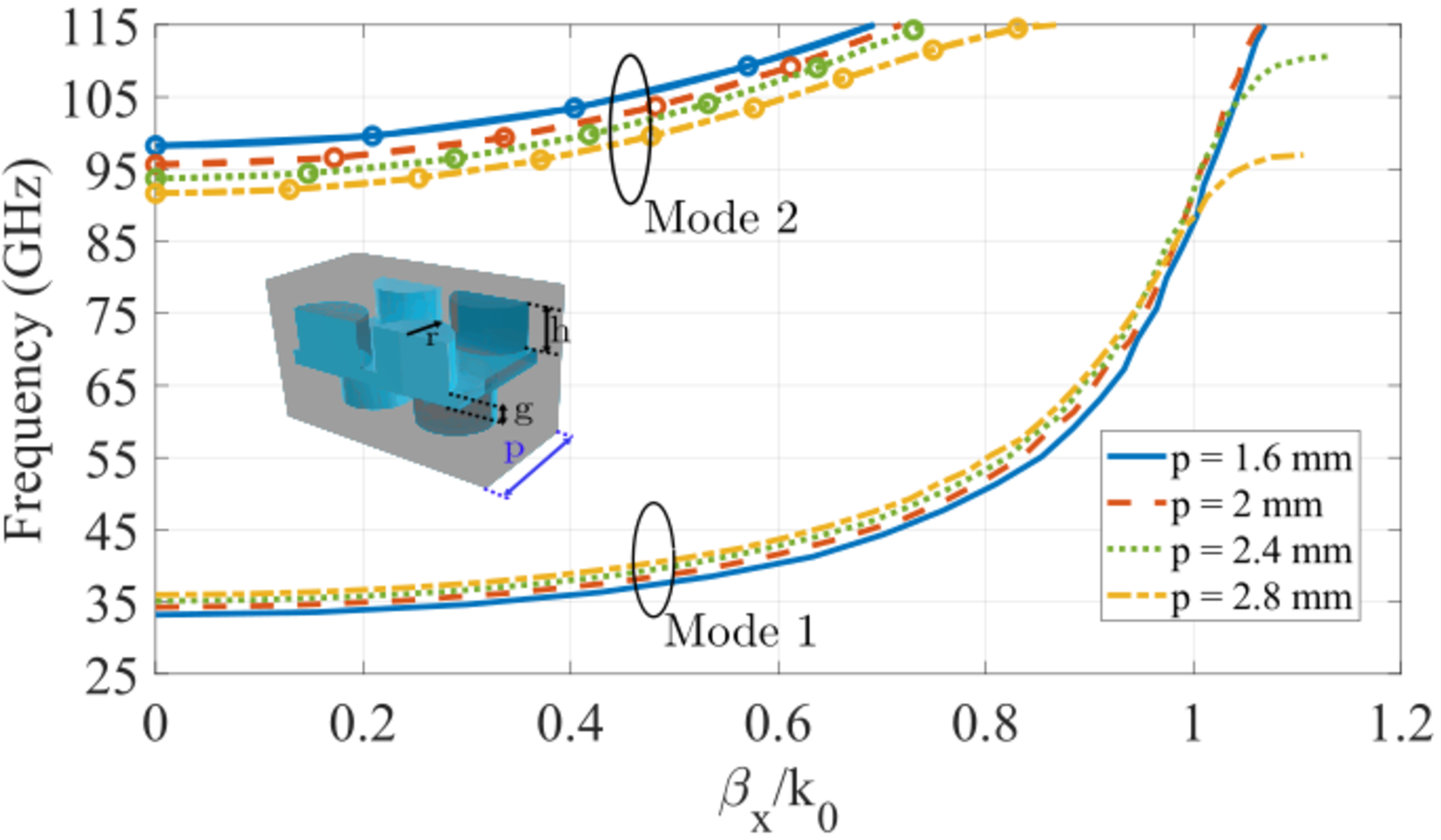}\label{figure6a}}
	\subfigure[]{\includegraphics[width= 0.415\textwidth]{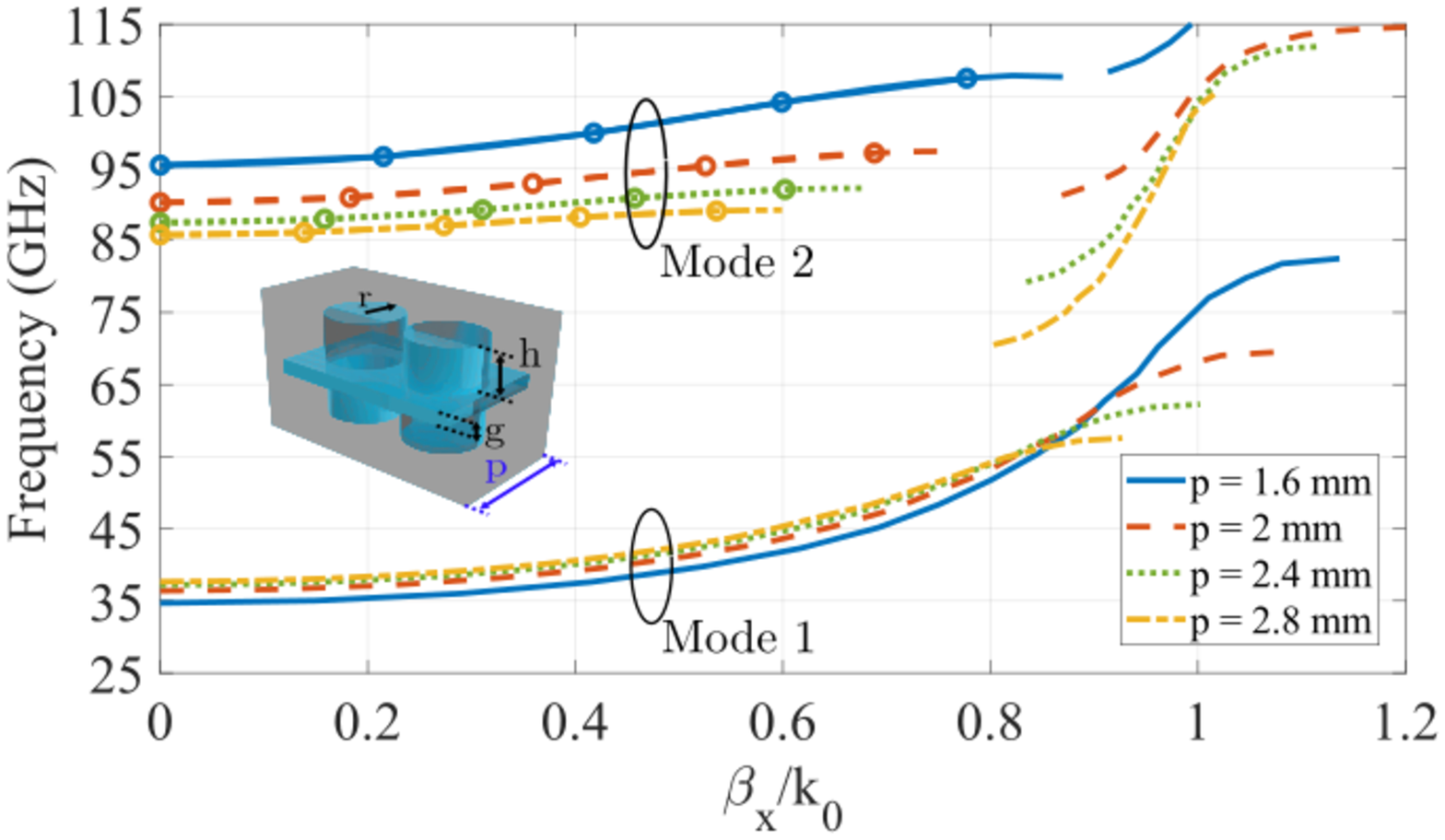}\label{figure6b}}
	
	\caption{Dispersion diagrams of the (a) Glide-symmetric and (b) Mirror-symmetric holey waveguides when varying the period of the unit cell. First and second propagating modes are displayed in the figure. The geometrical parameters of the unit cells are: \textit{g} = 0.3 mm, \textit{r} = 0.75 mm and \textit{h} = 1 mm.}
	\label{figure6}
\end{figure}

The effect of the period \textit{p} on the glide-symmetric and mirror-symmetric holey waveguides is shown in Fig. \ref{figure6}. Note that phase constant $\beta_x$ is normalized to $k_{0}$ in Fig. 6 to take into account that the period is varying. For both configurations, as the period increases, the holes are more distant and thus, their interaction decreases. For instance, the consequence in the mirror-symmetric case is the narrowing of the stopband region. Conversely, reducing the period, the stopband widens and displaces upwards in frequency.\\

\begin{figure}[t]
	\centering
	\subfigure[]{\includegraphics[width= 0.415\textwidth]{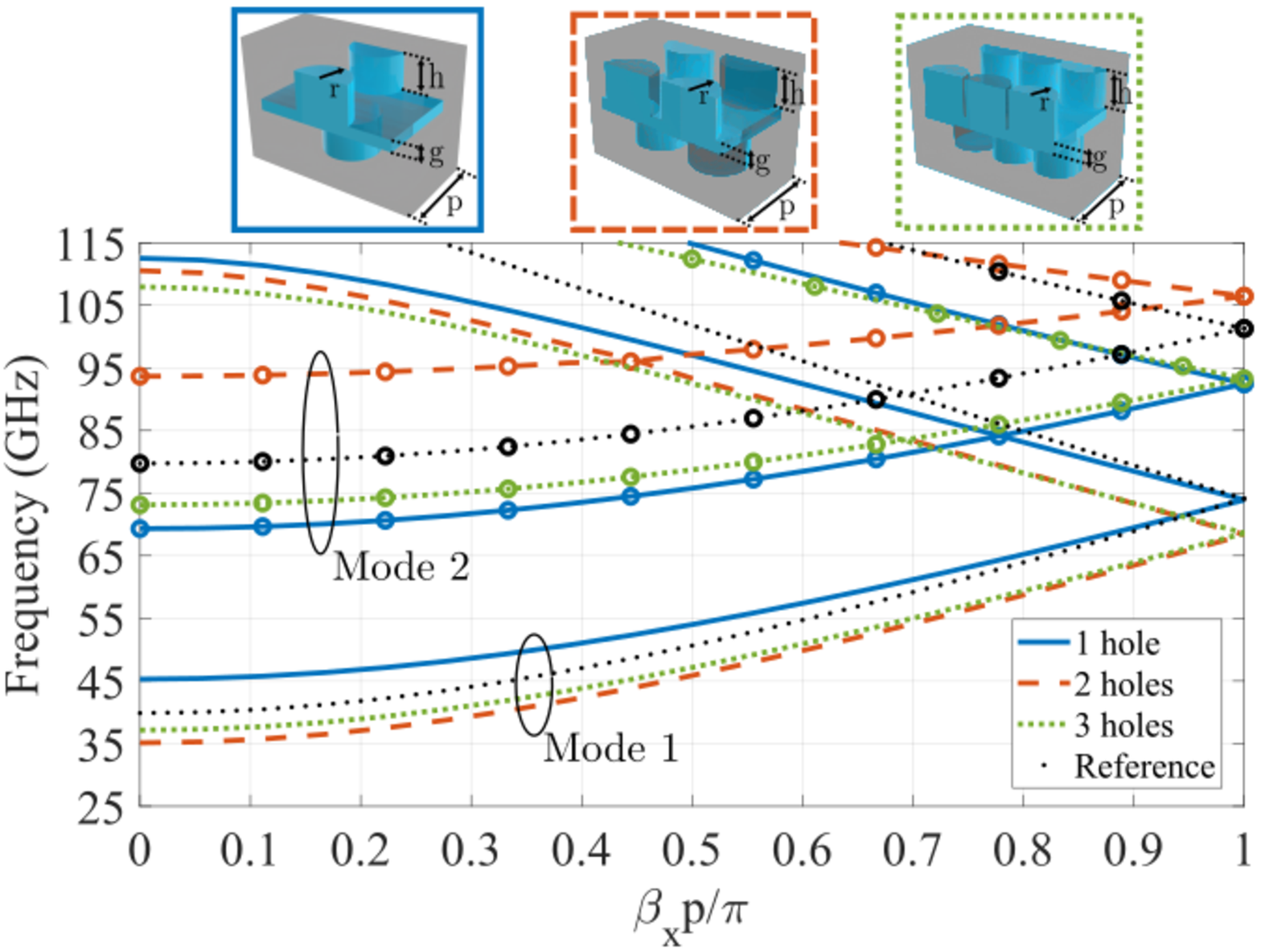}\label{figure7a}}
	\hspace{1mm}
	\subfigure[]{\includegraphics[width= 0.415\textwidth]{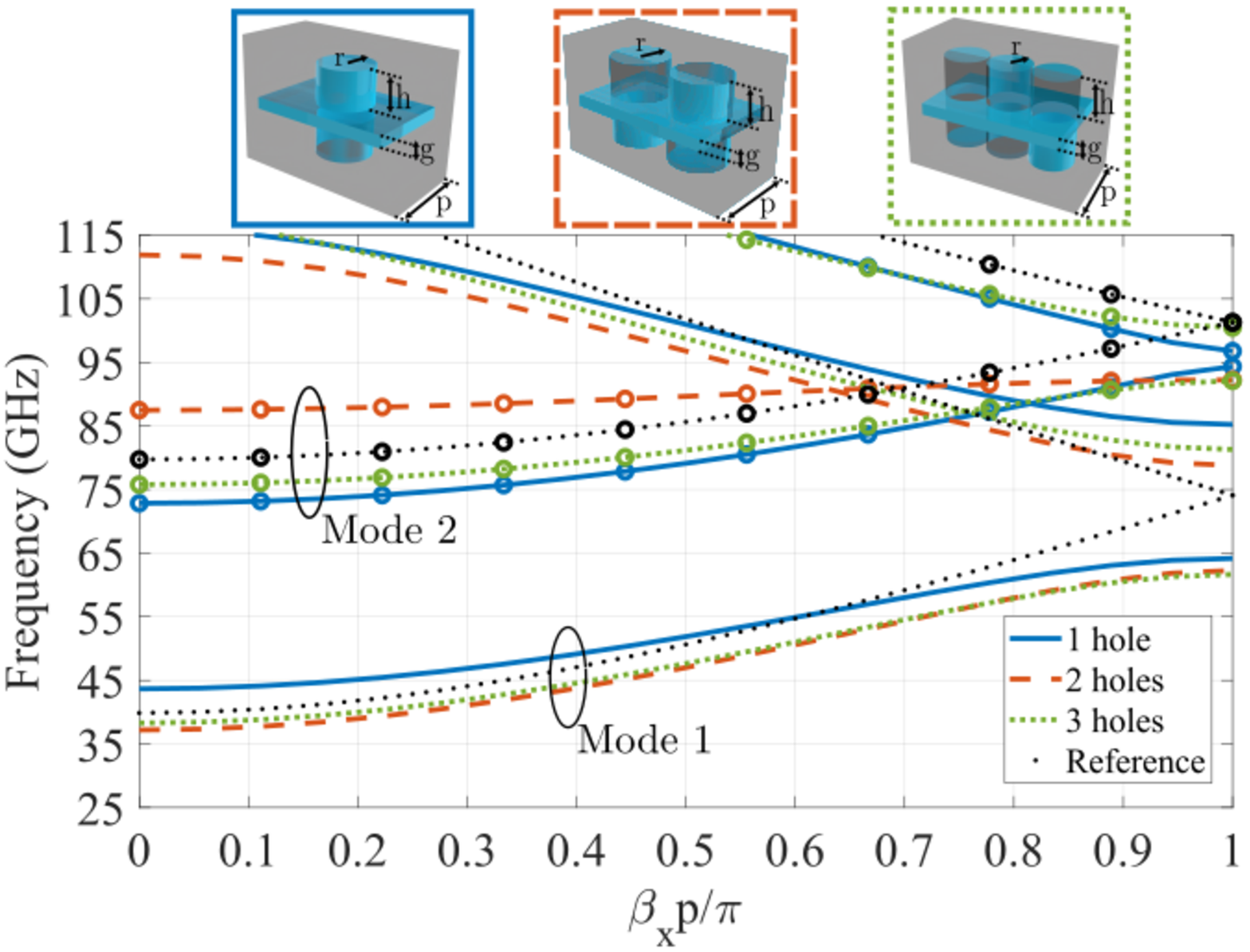}\label{figure7b}}
	\hspace{1mm}
	\subfigure[]{\includegraphics[width= 0.47\textwidth]{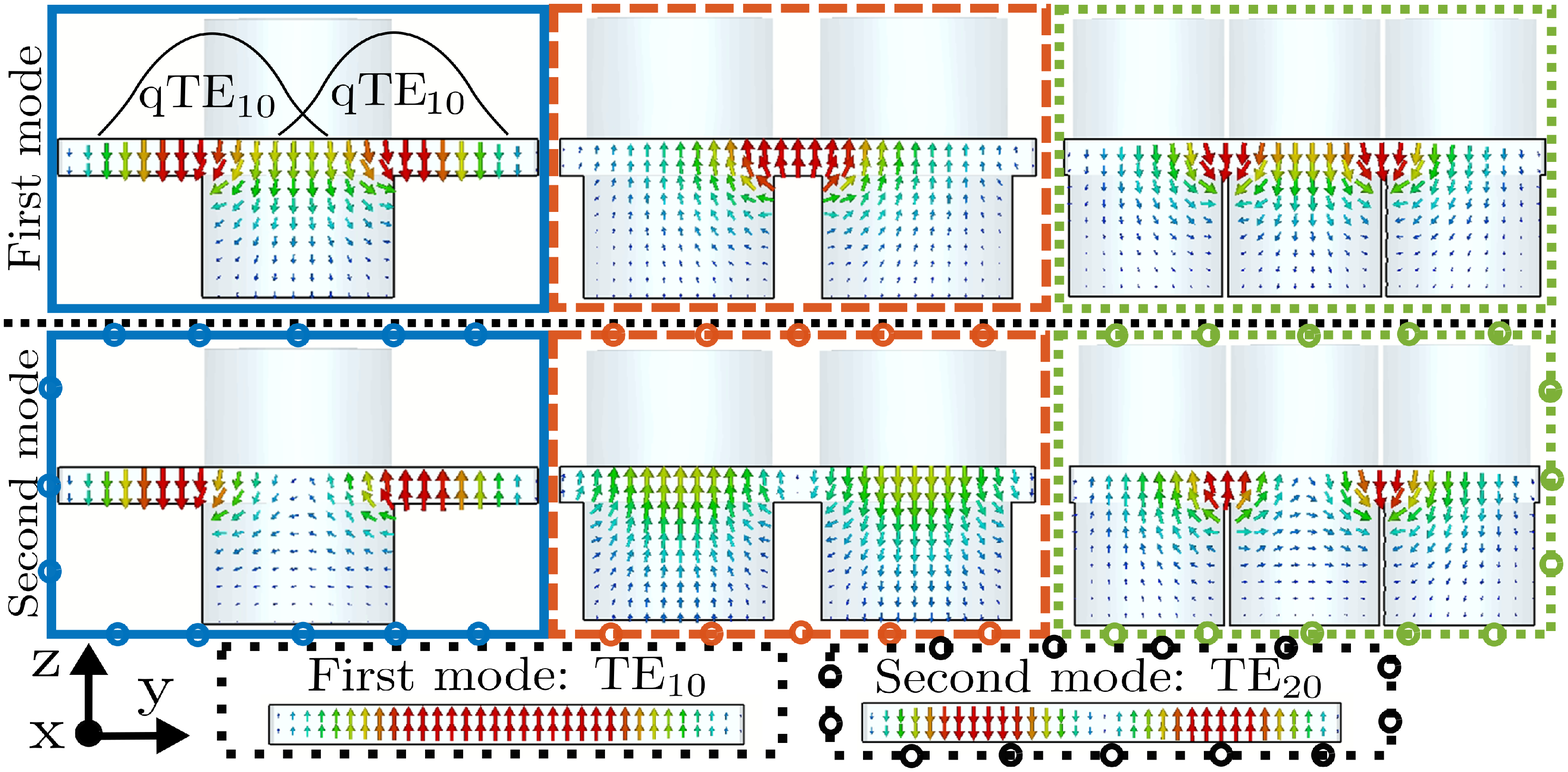}\label{figure7c}}
	
	\caption{Dispersion diagrams when the number of holes is modified for (a) The glide-symmetric and (b) Mirror-symmetric holey unit cells. The geometrical parameters of the unit cells are: \textit{p} = 2.4 mm, \textit{g} = 0.3 mm, \textit{r} = 0.75 mm and \textit{h} = 1 mm. First and second propagating modes are displayed in the figure. (c) Electric field distributions for the propagating modes in the reference and glide-symmetric unit cells when modifying the number of holes.}
	\label{figure7}
\end{figure}

The last parameter under study is the number of upper and lower holes that define the unit cell. The influence of the number of holes on the dispersion diagram is illustrated in Figs. \ref{figure7a} and  \ref{figure7b} for the glide-symmetric and mirror-symmetric configurations, respectively. The cutoff frequencies of the first and second propagating modes vary according to the number of holes placed in the unit cell. Furthermore, both configurations present the same effects when the number of holes is varied with the difference of the appearance of stopbands in the mirror-symmetric unit cell.\\

The amplitude distributions of the different modes provide a physical rationale for understanding the cutoff frequencies for each unit cell case. Thus, Fig. \ref{figure7c} illustrates  the electric field distributions of the propagating modes in the glide-symmetric unit cell as the number of holes is varied. For the sake of conciseness, the electric field of the mirror-symmetric unit cell has not been shown. In the case of a unit cell with a single hole (blue lines), the cutoff frequencies of the first and second modes are above and below the reference waveguide modes, respectively. The presence of a hole in the middle of the waveguide makes the TE\textsubscript{10} splits in two quasi-TE\textsubscript{10} modes that form the first mode (see blue box of Fig. \ref{figure7c}). The effective width of the quasi-TE\textsubscript{10} mode is shorter than the width of the original TE\textsubscript{10} and thus, the cutoff frequency for the first mode increases regarding the reference waveguide.
The value of the cutoff frequency takes into account all the effective widths along the length of the unit cell since the size of the hole varies along the \textit{x}-direction due to its circular geometry. However, it should be remarked that the illustrated transversal cut is shown at the center of the unit cell, that is, where the size of the hole is maximum and the effective width is minimum.

The second mode for the single-hole unit cell has a similar configuration that the TE\textsubscript{20} of the reference waveguide. A lower cutoff frequency compared to the reference waveguide is obtained due to the increase of the separation in $y$-direction between amplitude maxima of the second mode, which means the increase of the effective width for the mode. If the radius of the hole is progressively smaller, the cutoff frequency of the second mode increases turning into the TE\textsubscript{20} mode in the reference waveguide.

In the case of the unit cell with two holes in glide-symmetric configuration, the cutoff frequencies of the first and second modes are below and above the reference waveguide modes, respectively. Observing the electric field distributions of the modes, they resemble to the existing modes in a single ridged waveguide \cite{RWGs}. Comparing the mirror-symmetric and glide-symmetric unit cells, the differences in the cutoff frequencies are caused by the existence of transversal cuts without presence of holes, in the same manner that it occurs for the single-hole unit cell.

Lastly, the unit cell with three holes is analyzed. In this case, the electric fields for the first and second modes are similar to those observed in a double-ridged waveguide section \cite{DRWGs}. In addition, it is evidenced in Figs. \ref{figure7a} and  \ref{figure7b} that the phase constant for the first mode in the two-hole and three-hole unit cell are quite similar. This fact can be explained by the similarity between their electric field distributions (see red and green boxes in Fig. \ref{figure7c}).

\subsection{Comparison between holey and pin-loaded unit cells}

\begin{figure}[t!]
	\centering
	\subfigure[]{\includegraphics[width= 0.48\textwidth]{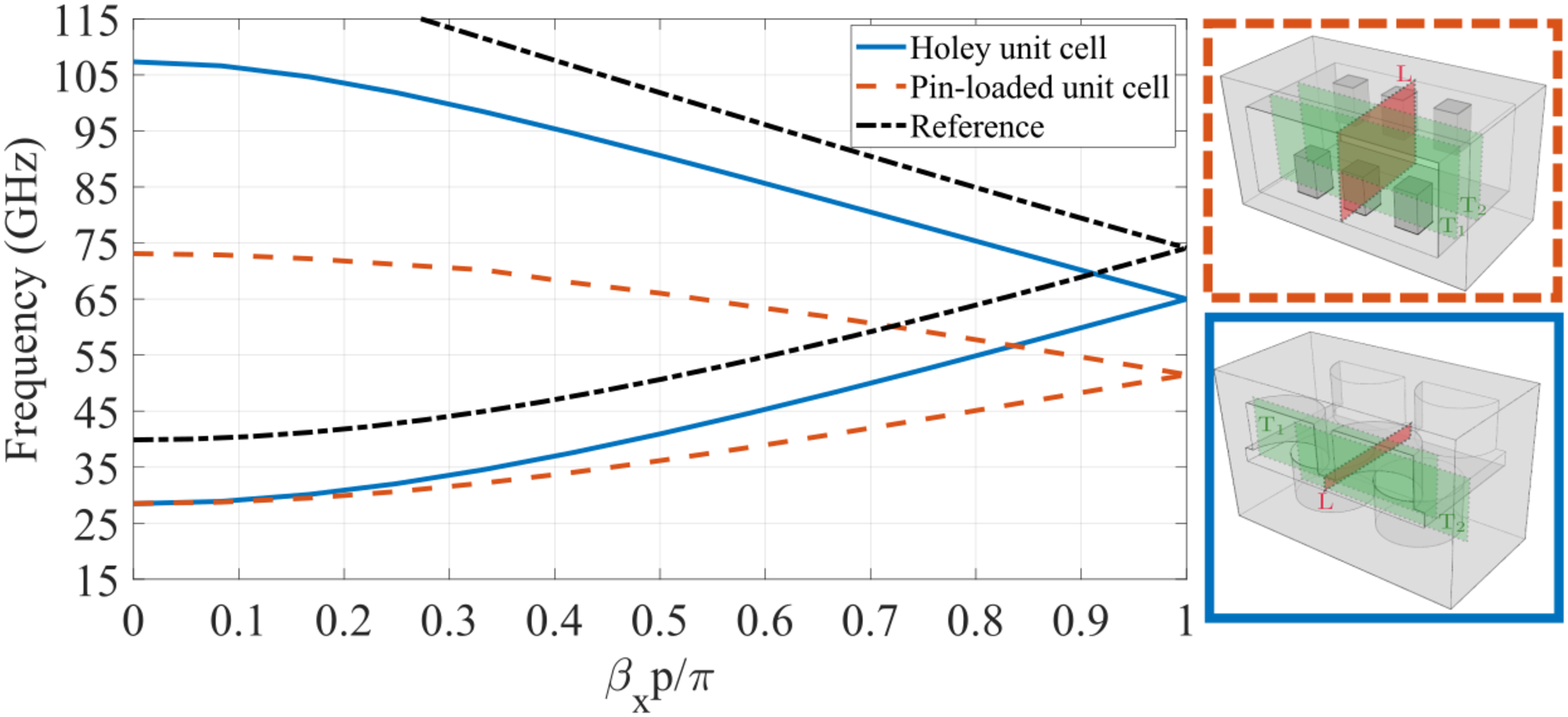}\label{figure8a}}
	\hspace{1mm}
	\subfigure[]{\includegraphics[width= 0.48\textwidth]{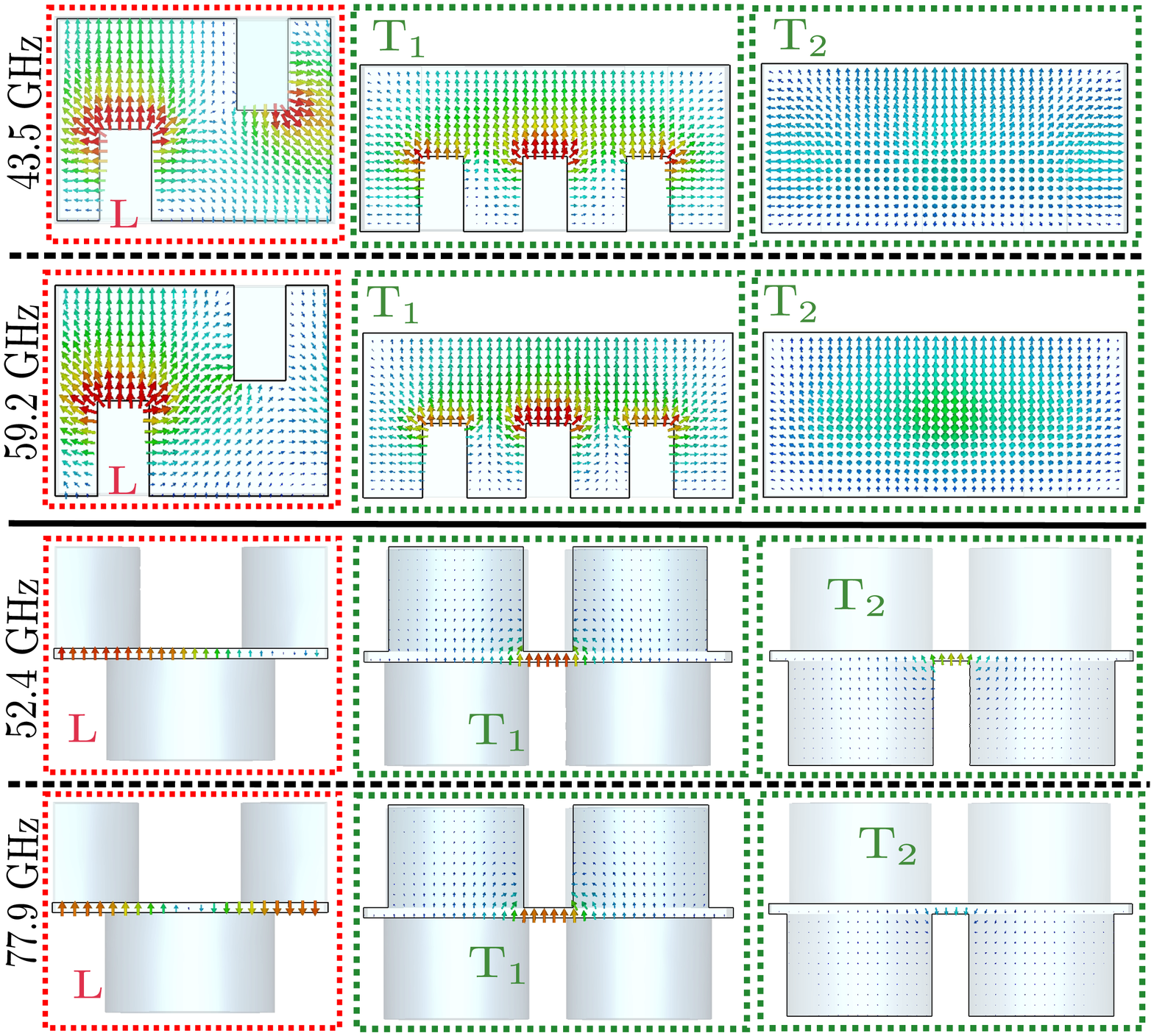}\label{figure8b}}

	\caption{Comparison between holey and pin loaded unit cells. (a) Dispersion diagrams of the first propagating mode and (b) Electric field distributions in different cut views at the same phase difference ($\beta$\textsubscript{x}p/$\pi$ $=$ 0.75) between limits of the unit cell (\textit{p} $=$ 2.4 mm). The first propagating mode is displayed in the figure.}
	\label{figure8}
\end{figure}

A dispersion comparison between the holey waveguide unit cell and its counterpart, the pin-loaded unit cell, is carried out in this subsection. The dispersion properties of the structure with pins in waveguide were analyzed in our previous work \cite{PinPhaseShifter_2}. The results show the beneficial effect of the glide-symmetric structure with pins for the design of waveguide phase shifters. Since both holey and pin-loaded waveguides can tune the propagation constant depending on the dimensions of holes and pins, respectively, it is of interest to analyze the advantages and disadvantages of their use. In Fig. \ref{figure8a}, the dispersion diagram of the fundamental mode in a glide-symmetric configuration is illustrated. In order to make a fair comparison, the geometrical parameters of both unit cells have been adapted so the same cutoff frequency with the same unit cell period \textit{p} and the same waveguide width (WR15 waveguide size) is obtained. Thus, the height of the pins has been set to 0.85 mm. It is observed that the pin-loaded waveguide is more frequency-dispersive compared to the holey waveguide. That is, the dispersion curve of the holey waveguide maintains approximately parallel to the reference curve in a wider bandwidth.

To get a physical insight on the behavior of both holey and structure with pins configurations, the electric field distribution at different sections of the unit cells is analyzed in Fig. \ref{figure8b}. For the computation, a fixed phase constant ($\beta$\textsubscript{x}p/$\pi$ $=$ 0.75) has been selected. As the mode is propagating in the unit cells, the pin-loaded design has more noticeable differences in the boundary conditions between transversal sections. The electric field in the transversal cut \textit{T\textsubscript{1}} has different boundary conditions due to the presence of pins compared to the ones in \textit{T\textsubscript{2}} where there is no structure with pins. This inhomogeneity in the structure along different transversal sections is not observed for the holey unit cell where the transversal sections are very similar among them approaching to a single-ridged waveguide section. This fact provokes that glide-symmetric holey waveguides are low-dispersive over a wide range of frequency in contrast with the dispersive behavior of the pin-loaded waveguide. This enhanced performance can be exploited for the design of wideband and low-dispersive components with fine tunability of their dispersive properties, as it was depicted in the previous subsection. In addition, the reduced size of pins as the frequency increases implies in general a higher manufacturing complexity compared to the implementation of holes.

\subsection{Stopband Analysis}

In this subsection, we study the filtering capabilities of the mirror-symmetric and broken glide-symmetric holey structures. We analyze the effect of the geometrical parameters that define the holey unit cells on the stopband region. The Eigenmode solver of \textit{CST} is not able to compute the attenuation in periodic structures, namely the real part $\alpha$ of the propagation constant $\gamma=\alpha +j\beta$. To solve this problem, we perform a Bloch analysis, similar to \cite{bloch_analysis1, bloch_analysis2}, based on the multi-modal transfer-matrix method detailed in the Appendix. The multi-modal transfer matrix \textbf{T} is extracted from the generalized scattering matrix \textbf{S} of a single unit  cell \cite{matrices1, matrices2}. The \textbf{S} matrix is computed via full-wave simulations with the frequency solver (tetrahedral mesh) of \textit{CST}.

\begin{figure}[t]
	\centering
	\subfigure[]{\includegraphics[width= 0.45\textwidth]{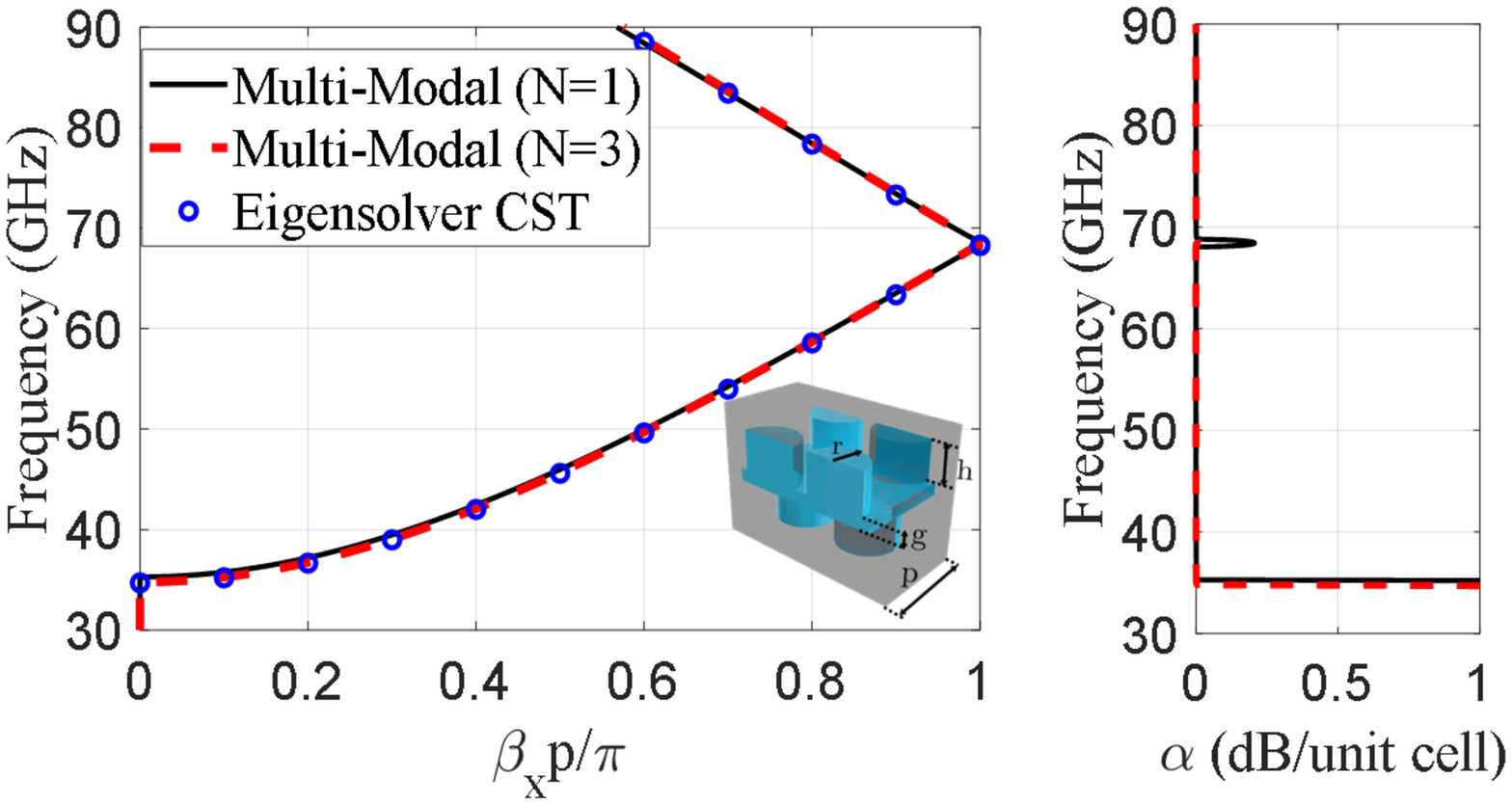}\label{figure9a}}
	\subfigure[]{\includegraphics[width= 0.45\textwidth]{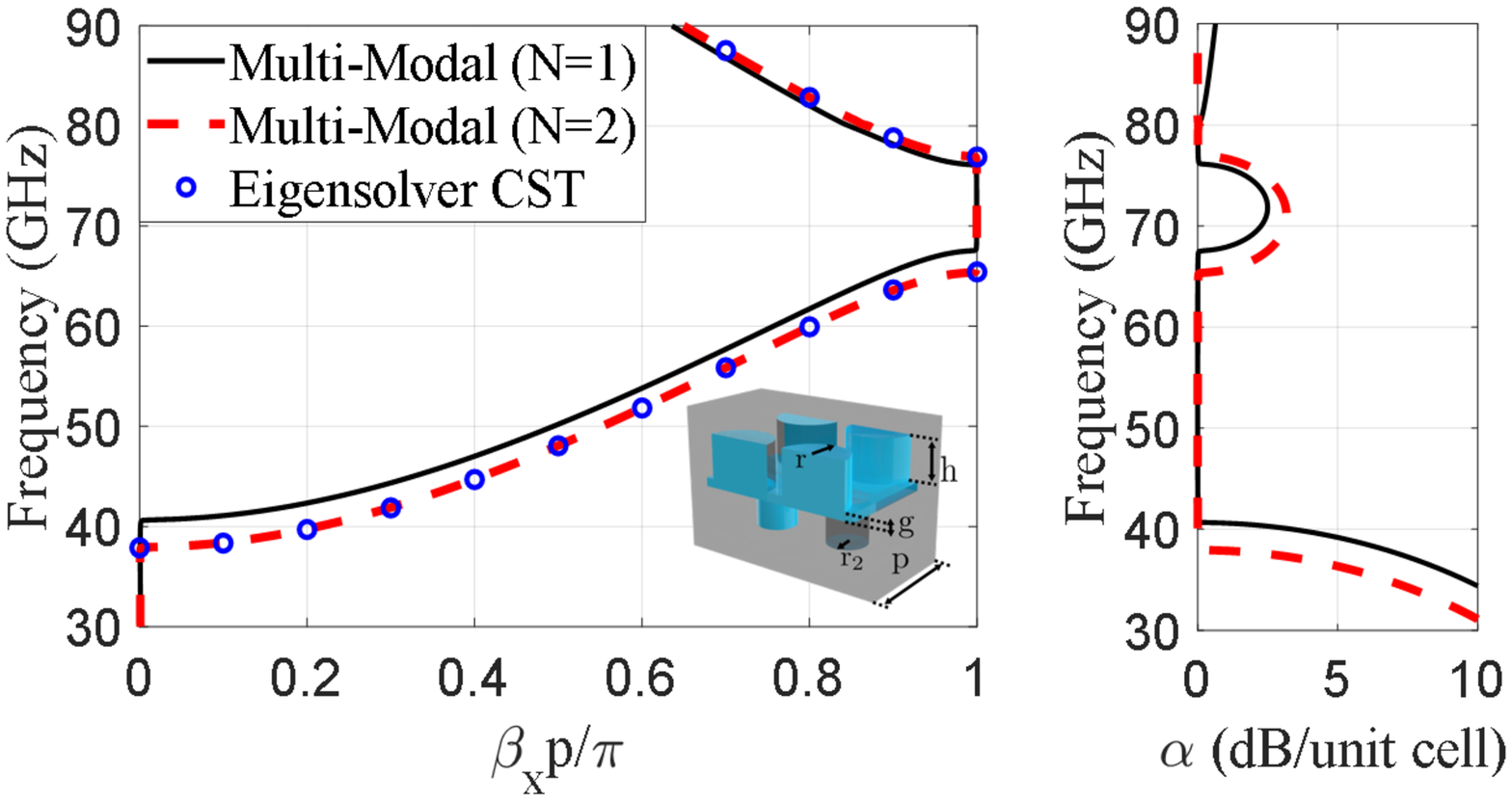}\label{figure9b}}
	\subfigure[]{\includegraphics[width= 0.45\textwidth]{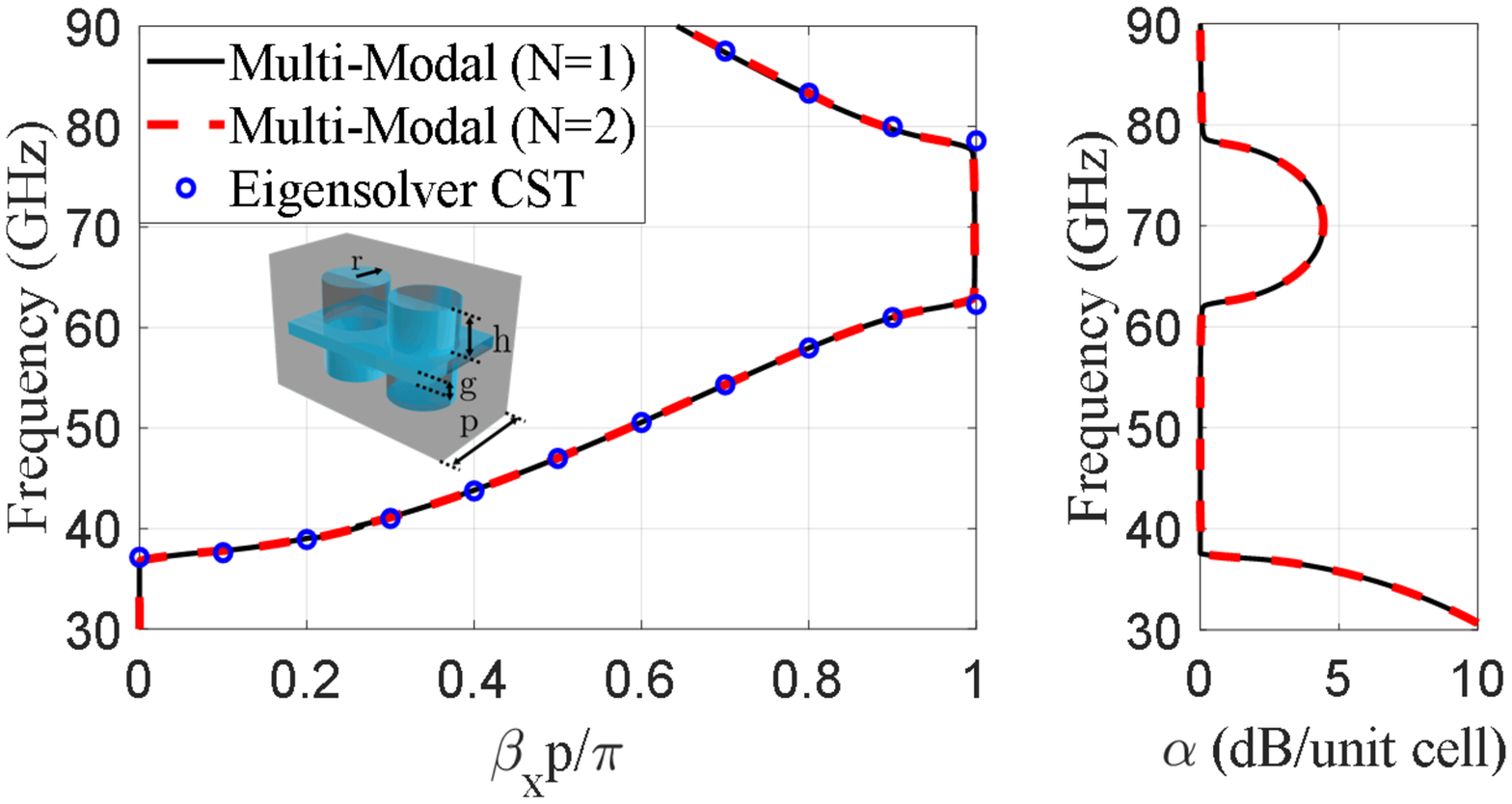}\label{figure9c}}
	\caption{Dispersion diagrams of the (a) Glide-symmetric, (b) Broken glide-symmetric ($r \neq r_2$) and (c) Mirror-symmetric unit cells with two holes. Attenuation ($\alpha$) and phase ($\beta$\textsubscript{x}) constants are shown in the figure. The geometrical parameters of the unit cells are: $r=0.72$ mm, $r_2=0.2$ mm, $g=0.25$ mm, $h=1$ mm, and $p=2.4$ mm.}
	\label{figure9}
\end{figure}

From the different configurations presented in Fig. \ref{figure9}, the most demanding for the multi-modal method are the glide-symmetric structures, due to the strong coupling of higher-order Floquet harmonics \cite{bloch_analysis2, coupling1, symmetry_Paco, coupling2}.  Thus, mirror-symmetric and broken glide-symmetric structures  will demand, in general, a lower number of modes $N$. During the computation of the results, it is observed that the choice of modes that possess a maximum of electric field in the center of the waveguide (area between holes) favours the convergence of results.  It is found that $N=3$ modes are needed in the glide-symmetric holey configuration. See that the undesired stopband disappears and the results converge. $N=2$ modes are used in the computation of broken glide-symmetric structures. Only one mode ($N=1$) is needed in the mirror-symmetric configuration due to the lower higher-order coupling between modes.

As previously discussed in Section II.A, the glide-symmetric structure suppresses the stopband between first and second modes, so the fundamental mode propagates in the entire WR15 frequency range. This fact can be appreciated in Fig. \ref{figure9a}. In order to create a stopband in the glide-symmetric structure and exploit its filtering properties, the glide symmetry must be conveniently broken. This can be achieved by using holes of different radii and heights, as previously discussed in \cite{GlideSymm_holes3}. Alternatively, work in \cite{GlideSymm_holes4} presents a holey unit cell with braided glide symmetry that provides a high attenuation value. In this work, different configurations of the holes are depicted in detail to produce stopbands in holey unit cells. As an example, Fig. \ref{figure9b} shows the creation of a stopband from 65 GHz to 75 GHz after breaking glide symmetry by modifying the radius $r_2$ of the intermediate bottom holes. On the other hand, the mirror-symmetric configuration shows an inherent stopband between the first and second modes, as shown in Fig. \ref{figure9c}, located approximately at the upper part of the WR15 frequency range. Good agreement is observed in the computation of the phase constant $\beta$\textsubscript{x} between the multi-modal method and the Eigenmode solver of \textit{CST}. Note that no comparison is shown for the attenuation constant $\alpha$, since the Eigenmode solver of \textit{CST} cannot compute it.

\begin{figure}[t]
	\centering
	\subfigure{\includegraphics[width= 0.47\textwidth]{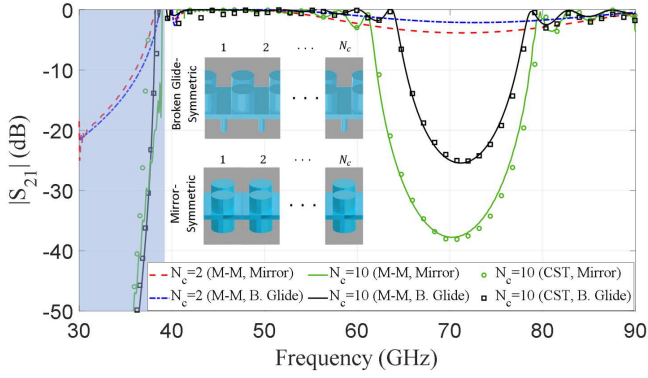}\label{glide_alpha_beta_mirror_fabricado}}
	\caption{$|S_{21}|$ parameter of finite mirror-symmetric and broken glide-symmetric structures as a function of the number of concatenated unit cells $N_c$. The cutoff range of the waveguide is shadowed in blue. In the legend, M-M states for the multi-modal technique. The geometrical parameters of the unit cell are: $r=0.72$ mm, $g=0.25$ mm, $h=1$ mm, and $p=2.4$ mm.}
	\label{figure10}
\end{figure}

Finite number of concatenated unit cells can be also studied with the use of the multi-modal method by simply cascading the transfer matrix. This is of interest for the design of filters made of periodic arrangements. Simulation time and computational efforts are saved, since simulating a large number of concatenated cells in a commercial software like \textit{CST} is time consuming. Fig. \ref{figure10} presents the $|S_{21}|$ parameter of the broken glide-symmetric (labeled as ``B. Glide") and mirror-symmetric (labeled as ``Mirror") structures as a function of the number of concatenated unit cells, $N_c$. The stopband centered at 70 GHz observed in Figs. \ref{figure9b} and \ref{figure9c} is also appreciated in Fig. \ref{figure10}. As shown in these figures, the mirror-symmetric configuration offers a higher attenuation constant in the rejection band compared to the broken glide-symmetric one. This will be discussed in detail later. Note that the maximum value of the attenuation constant $\alpha$ is slowly converging to the values evidenced in Fig. \ref{figure9}  as the number of concatenated unit cells $N_c$ increases. As an example, the maximum attenuation constant is  4.35 dB/unit cell for the mirror-symmetric structure with periodic holes, while in the finite mirror-symmetric structures: $\alpha=1.92$ dB/unit cell ($N_c=2$), and $\alpha=3.82$ dB/unit cell ($N_c=10$).  For validation purposes, a full-wave simulation in \textit{CST} of the complete structures with 10 unit cells is also presented in Fig. \ref{figure10} (green circles and black squares). As shown, there is a good agreement between \textit{CST} and the multi-modal approach.

\begin{figure}[h]
	\centering
	\subfigure[]{\includegraphics[width= 0.45\textwidth]{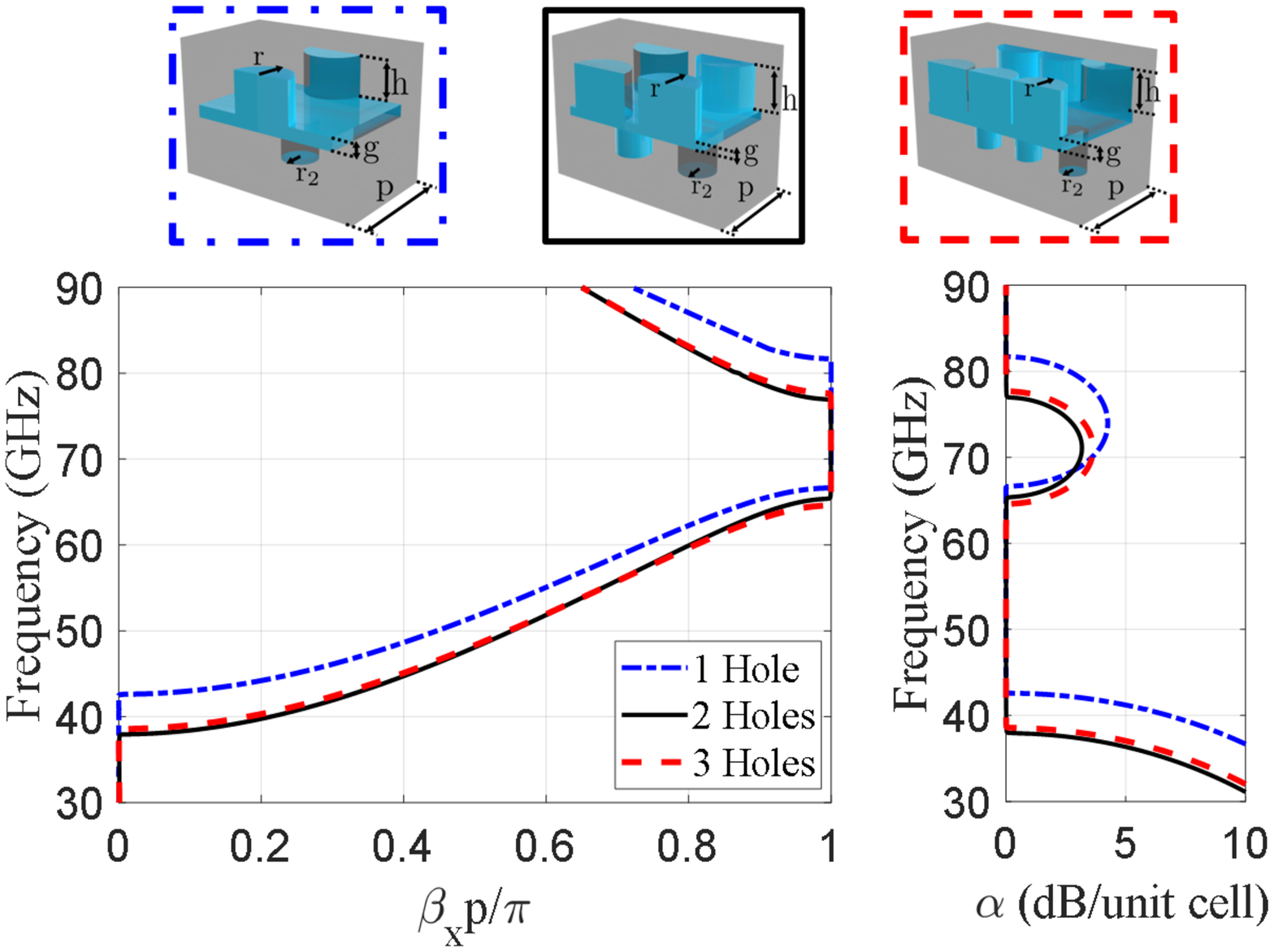}\label{figure11a}}
	\subfigure[]{\includegraphics[width= 0.45\textwidth]{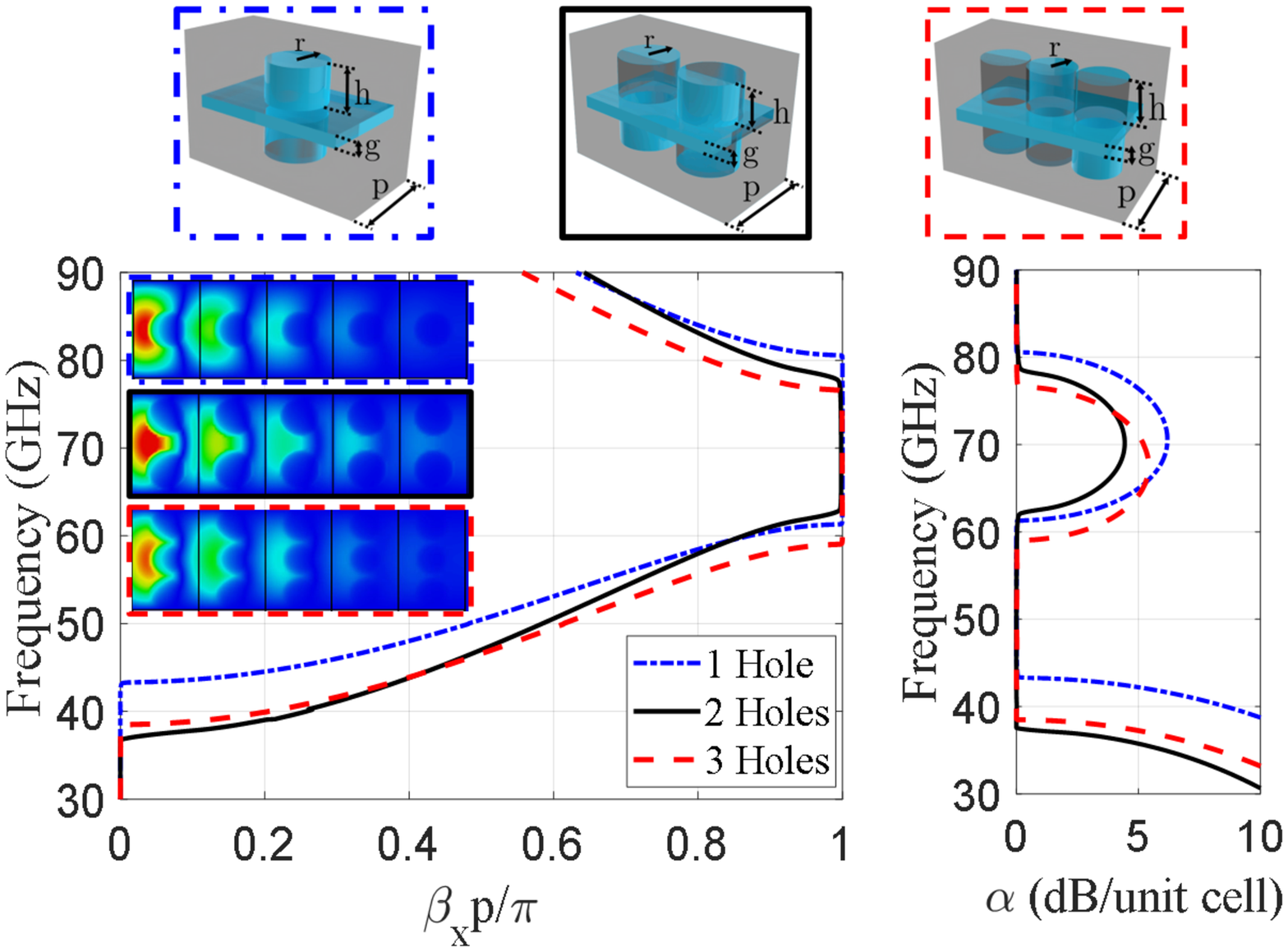}\label{figure11b}}
	\caption{Dispersion diagrams of the (a) Broken glide-symmetric and  (b) Mirror-symmetric structures as a function of the number of holes. Attenuation ($\alpha$) and phase ($\beta$\textsubscript{x}) constants are shown in the figure. The inset shows the electric field at the stopband region (70 GHz) for the three cases under study. The geometrical parameters of the unit cells are: $r_{1}=0.72$ mm (1 and 2 holes), $r=0.60$ mm (3 holes), $r_2=0.20$ mm, $g=0.25$ mm, $h=1$ mm, and $p=2.4$ mm.}
	\label{figure11}
\end{figure}

Fig. \ref{figure11} presents the influence on the dispersion diagram of the number of holes placed in the unit cell. Both broken glide-symmetric and mirror-symmetric configurations follow the same pattern. The attenuation constant is higher for the case of a single hole and lower for the case of two holes. Placing three holes is an intermediate case. This fact can be better understood by looking at the electric field at the stopband region (70 GHz) of several concatenated unit cells. The attenuation constant is greater in the case of a single hole because the maximum amplitude of the fundamental mode coincides with the center of the hole. Thus, in this case, the mode is more strongly perturbed than the other two cases. Conversely, $\alpha$ is lower in the case of placing two holes since the maximum of the electric field occupies precisely the undrilled region between these two holes and, as a consequence, the wave propagation is less perturbed.

\begin{figure}[t!]
	\centering
	\subfigure[]{\includegraphics[width= 0.45\textwidth]{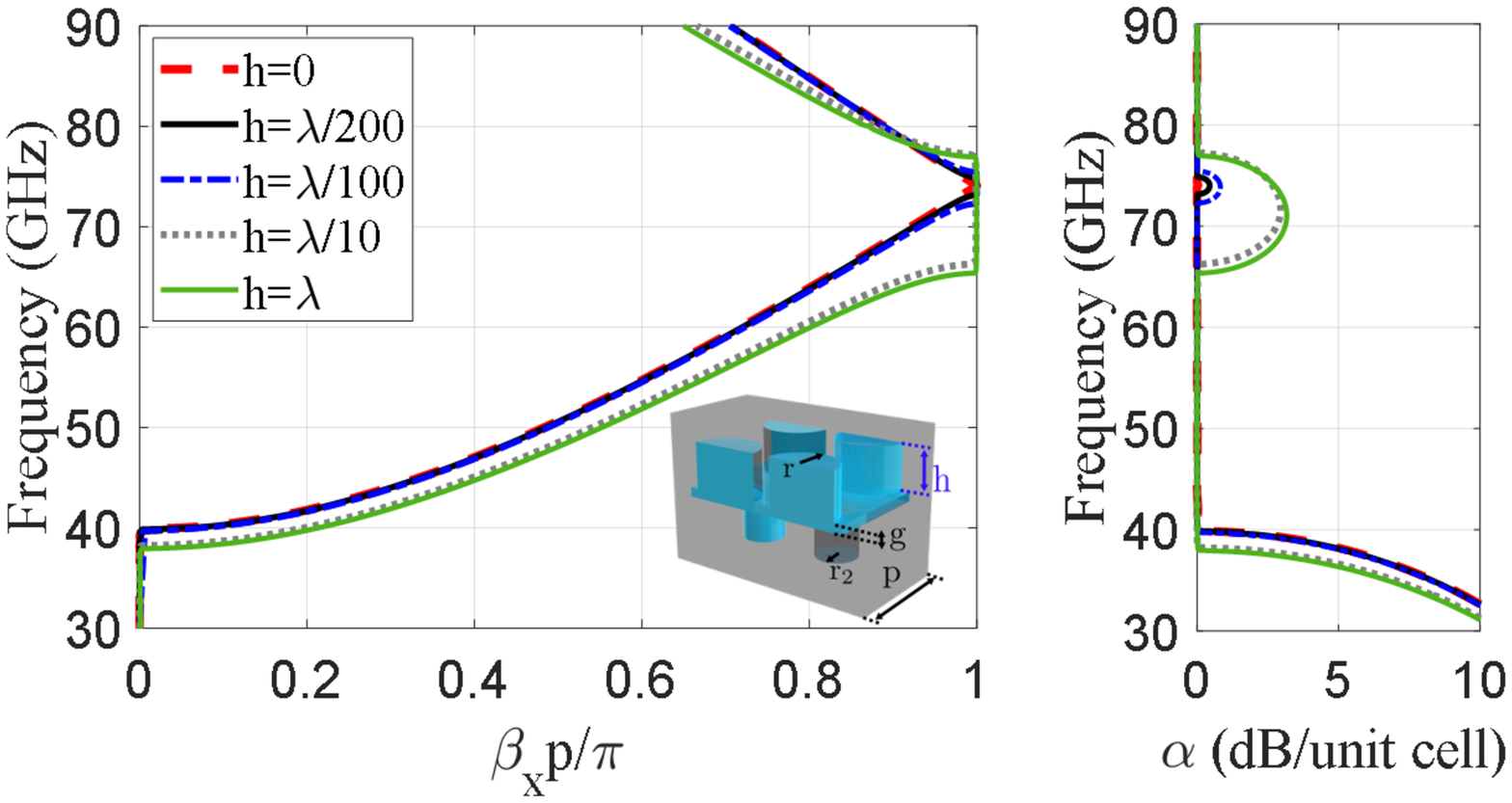}\label{figure12a}}
	\subfigure[]{\includegraphics[width= 0.45\textwidth]{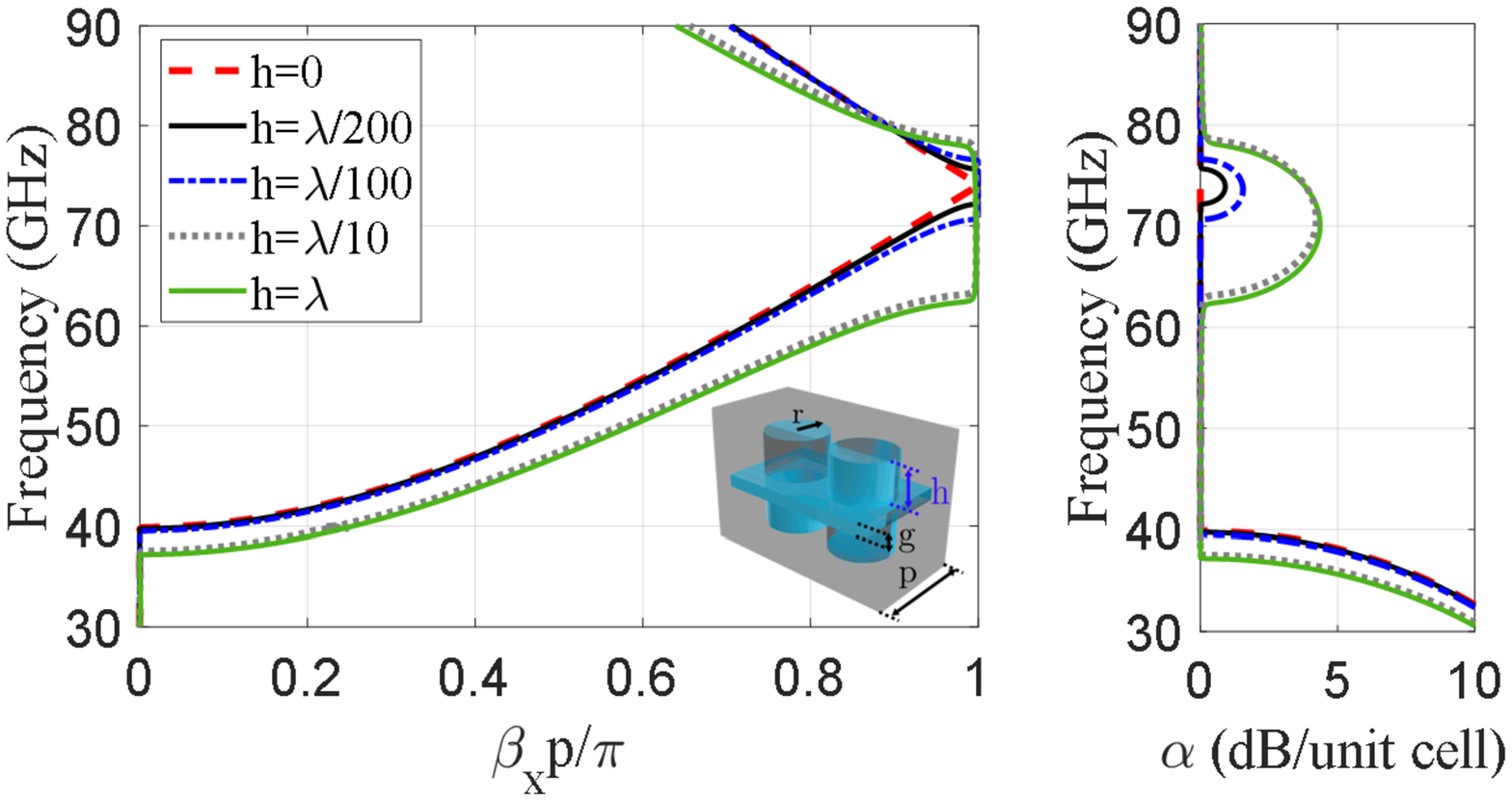}\label{figure12b}}
	\caption{Dispersion diagrams of the (a) Broken glide-symmetric ($r\neq r_2$) and (b) Mirror-symmetric unit cells with two holes as a function of the hole depth. Attenuation ($\alpha$) and phase ($\beta$\textsubscript{x}) constants are shown in the figure. The geometrical parameters of the unit cells are: $r_{1}=0.72$ mm, $r_2=0.20$ mm, $g=0.25$ mm, and $p=2.4$ mm.}
	\label{figure12}
\end{figure}

Fig. \ref{figure12} illustrates the influence of the hole depth $h$ on the attenuation and phase constants for the broken glide-symmetric and mirror-symmetric structures. In general, the deeper the hole is, the wider the stopband and the higher the attenuation constant are.  However, there is a limit value where increasing the hole depth has no longer influence in the dispersion diagram. Note that for $h=0$, the holey waveguide turns into a common hollow waveguide and the stopband closes. The parameter $h$ was expressed in terms of wavelength (at 70 GHz) to give a further insight on the physical mechanisms that rule the structures. From that, it can be appreciated that such electrically-small values as $h=\lambda/200$ ($21\, \mu$m) provoke noticeable attenuations in the stopband region: 0.42 dB/unit cell and 0.88 dB/unit cell for the broken glide-symmetric and mirror-symmetric configurations, respectively. Therefore, in waveguides with closely-spaced metal plates, any 
small protuberances repeated along the waveguide may cause noticeable losses due to the opening of narrow stopbands in the frequency range of the fundamental mode. These losses also exist for holey waveguides whose small protuberances are repeated in a non-periodic way. If the mirror-symmetric unit cell defines a holey waveguide but the locations and sizes of the holes vary among the unit cells, the produced losses will be variable and depend on the non-periodic configuration. However, its drop in transmission will be around the stopband that appears in the mirror-symmetric unit cell.

\begin{figure}[t!]
	\centering
	\subfigure[]{\includegraphics[width= 0.45\textwidth]{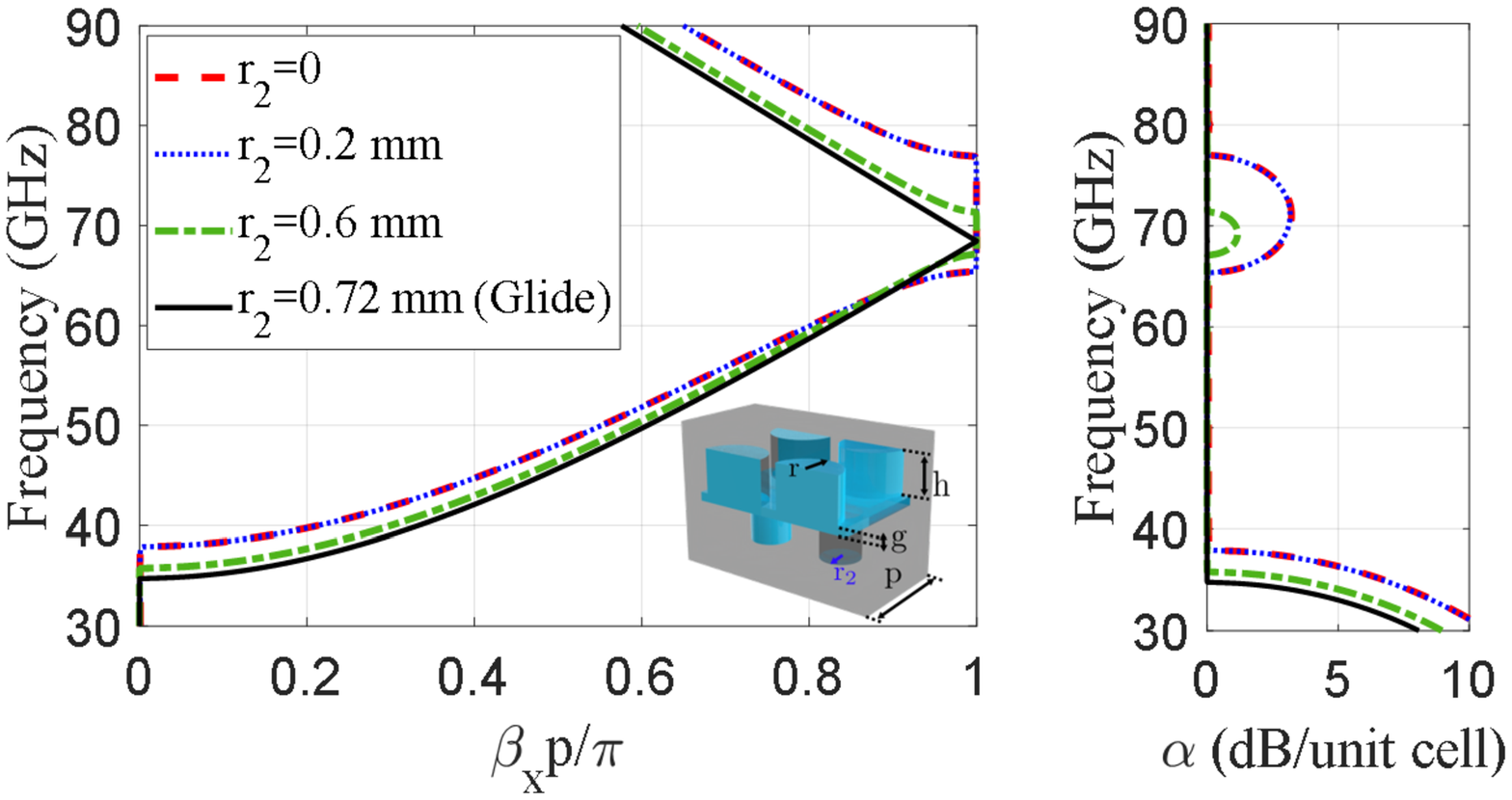}\label{figure13a}}
	\subfigure[]{\includegraphics[width= 0.5\textwidth]{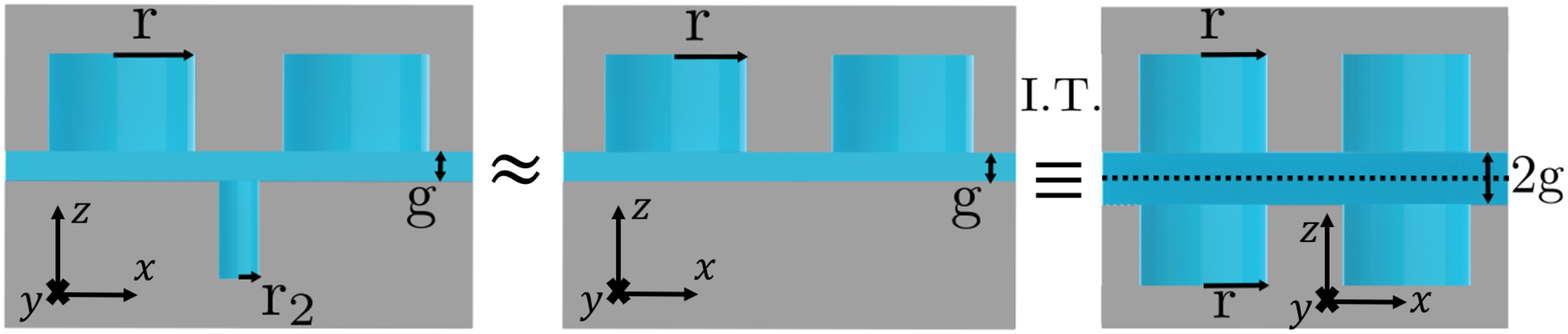}\label{figure13b}}
	\caption{(a) Dispersion diagram of the broken glide-symmetric ($r\neq r_2$) unit cell with two holes as a function of the radius $r_2$. Attenuation ($\alpha$) and phase ($\beta$\textsubscript{x}) constants are shown in the figure. The geometrical parameters of the unit cell are: $r_{1}=0.72$ mm, $g=0.25$ mm, $h=1$ mm, and $p=2.4$ mm. (b) Illustration of the performance of 1D broken glide-symmetric structures for small values of $r_2$.}
	\label{figure13}
\end{figure}

The radius of the intermediate holes $r_2$ determines the stopband region and the attenuation factor in the broken glide-symmetric configuration. Fig. \ref{figure13a} illustrates the dispersion diagram of the broken glide-symmetric structure with two holes when varying the radius $r_2$. It can be appreciated that smaller values of $r_2$ provoke wider stopbands and higher attenuation constants. In addition, note that the dispersion curves are quite similar in the cases $r=0$ (no intermediate holes) and $r=0.20$ mm. This fact allows us to explain why in Figs. \ref{figure11} and \ref{figure12} the mirror-symmetric structure presents a higher attenuation constant compared to the broken glide-symmetric one. The effect of the intermediate holes in the dispersion curves is negligible for small values of $r_2$. Therefore, the intermediate holes can be eliminated and the operation of the periodic structure remains the same, as shown in the schematic of Fig. \ref{figure13b}. By applying image theory, the bottom metallic plate of the waveguide (situated at a distance $g$) will create an image of the upper holes at a distance $2g$. That is, the broken glide-symmetric configuration for small values of $r_2$ is equivalent to a mirror-symmetric structure of double gap height $2g$, which offers a smaller attenuation constant compared to a mirror-symmetric structure of gap height $g$ (see Fig. \ref{figure5b}). It should be remarked that image theory is applicable because of the mirror symmetry of the unit cell.

\section{Applications: waveguide phase shifter and filter}

\begin{figure*}[t]
	\centering
	\subfigure[]{\includegraphics[width= 0.34\textwidth]{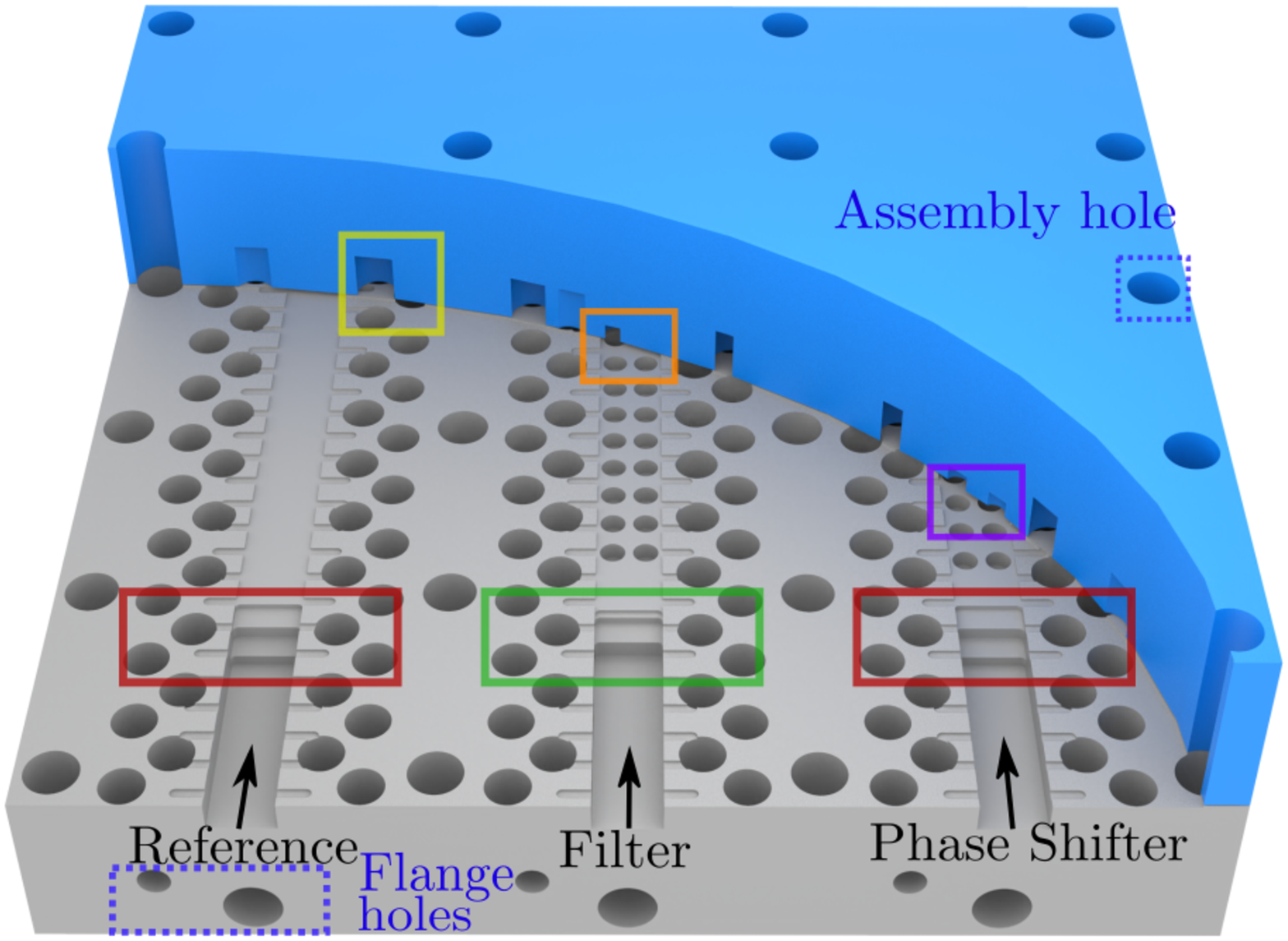}\label{figure14a}}
	\subfigure[]{\includegraphics[width= 0.32\textwidth]{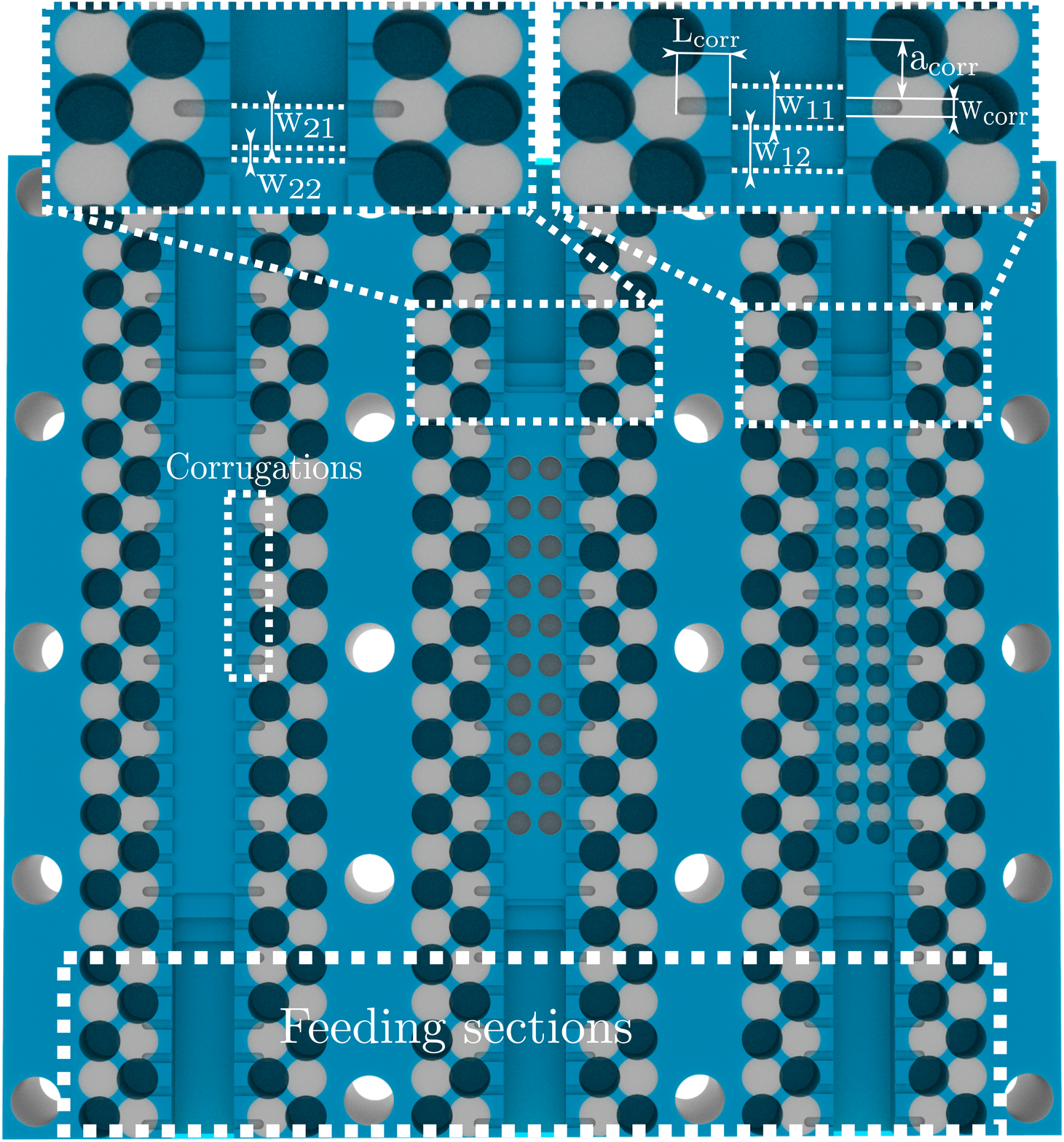}\label{figure14b}}
	\subfigure[]{\includegraphics[width= 0.32\textwidth]{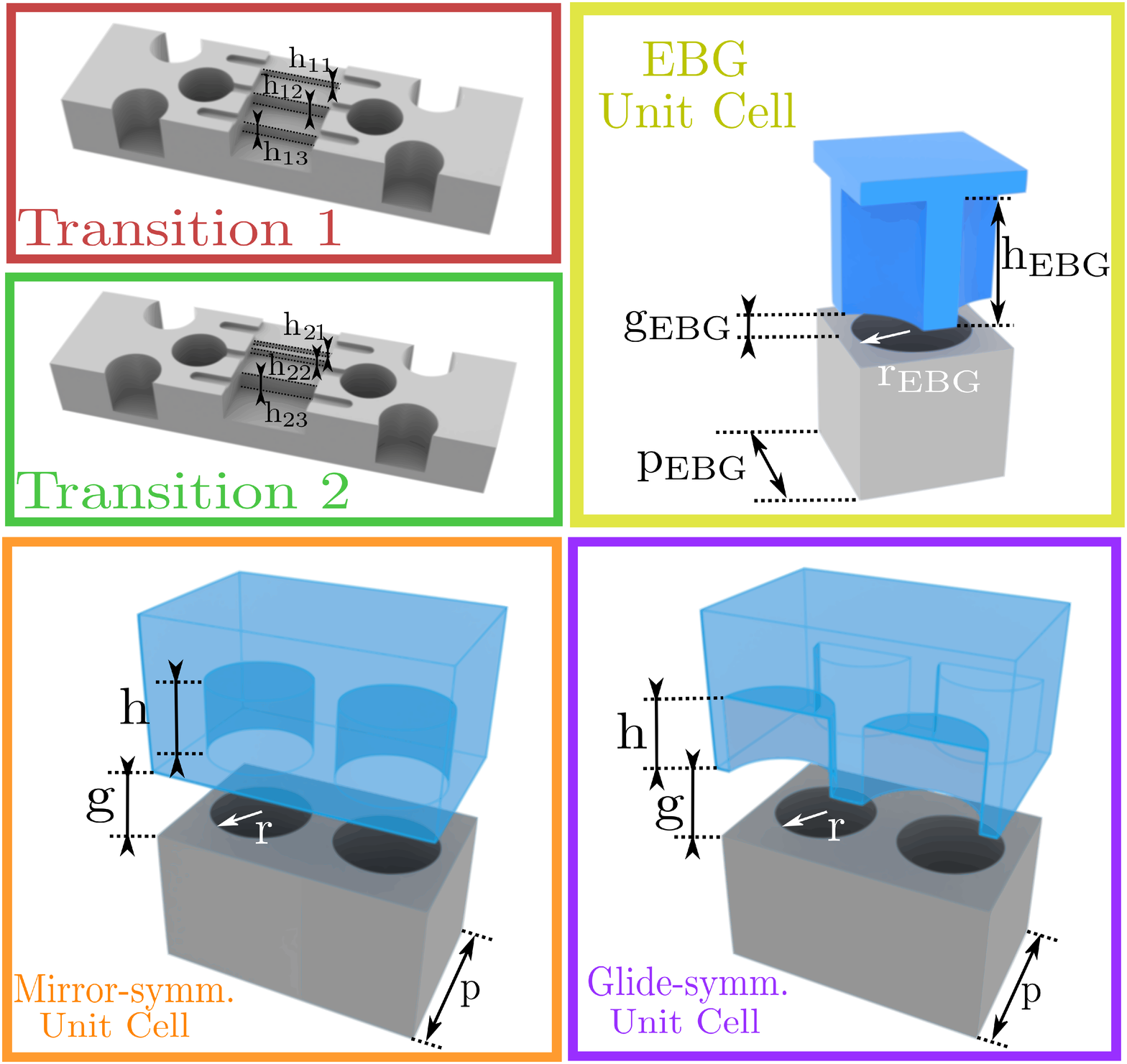}\label{figure14c}}
	
	\caption{Design of the prototype composed by the reference waveguide, filter and phase shifter: (a) 3D view, (b) Planar view of the forming layers and (c) Relevant geometries that composed the design. The dimensions are: $w_{11}=1.5$ mm, $w_{12}=1.5$ mm, $w_{21}=1.5$ mm, $w_{22}=0.4$ mm, $L_\mathrm{corr}=1.8$ mm, $a_\mathrm{corr}=2$ mm, $w_\mathrm{corr}=0.6$ mm, $h_{11}=0.28$ mm, $h_{12}=0.7$ mm, $h_{13}=0.6$ mm, $h_{21}=0.18$ mm, $h_{22}=0.4$ mm, $h_{23}=1$ mm, $g_\mathrm{EBG}=0.05$ mm, $r_\mathrm{EBG}= 1.25$ mm, $p_\mathrm{EBG}=3.22$ mm, $h_\mathrm{EBG}=3$ mm, $g=0.3$ mm, $r=0.72$ mm, $p=2.4$ mm and $h=1$ mm.}
	\label{Figura14}
\end{figure*}

Based on  the results obtained in the previous section for the holey unit cells, two proof-of-concept prototypes have been fabricated. The selected designs are a wideband low-dispersive phase shifter and a filter in order to validate the performance of the glide-symmetric and mirror-symmetric unit cells, respectively. Their dispersion diagrams correspond to Figs. \ref{figure3a} and \ref{figure3b} (green lines) with a slight modification in the size of the hole radius. The mirror-symmetric unit cell is selected in the design of the filter because it produces greater attenuation in the stopband compared to the broken glide-symmetric unit cell. A reference WR15 waveguide is also included to compare the performance of the phase shifter and the filter. The three designs have been implemented in the same metallic pieces, as it is illustrated in Fig. \ref{figure14a}. Planar view and key geometries are depicted in Figs. \ref{figure14b} and \ref{figure14c}.  The phase shifter and the filter are composed by a concatenation of ten glide-symmetric and mirror-symmetric holey unit cells, respectively. The waveguide has been split in two different parts (blue and grey pieces) in order to ease the manufacturing process. The operation frequency range is between 50 to 75 GHz. In order to prevent field leakage throughout the air gap due to an imperfect assembly, we make use of the gap-waveguide technology \cite{GlideSymm_holes1}. Thus, glide-symmetric EBG (electromagnetic bandgap) holes, marked in yellow in Fig. \ref{figure14a}, are placed on both sides of the reference waveguide, filter and phase shifter. As the width of the holey waveguide unit cells is \textit{w} = 3.76 mm, the feeding sections are waveguides with the width and the height corresponding to the WR15 waveguide to have appropriate flange connections in the setup. Nevertheless, waveguide transitions between feeding sections and the designs are required since the height of the waveguide in the unit cells are set to \textit{g} = 0.3 mm. Two different transitions are implemented for the phase shifter and filter, because they have different input impedances. The transition implemented for the reference WR15 waveguide is the same as for the phase shifter, in order to avoid any additional phase shift discrepancy introduced by the feeder.

The dimensions of glide-symmetric EBG holes are $p_\mathrm{EBG}=3.22$ mm, $r_\mathrm{EBG}= 1.25$ mm, $h_\mathrm{EBG}=3$ mm, and $g_\mathrm{EBG}=0.05$ mm. In this manner, the applied gap-waveguide technology can operate from 45 GHz to 85 GHz since this is the location of the bandgap in the EBG structure. The parameter $g_\mathrm{EBG}$ indicates the gap between layers, so the dimension of this parameter is selected to be the value at the worst case. In general, smaller gap heights provide wider stopbands.

Despite of the presence of glide-symmmetric EBG holes to prevent leakage throughout the gap, field resonances may exist at certain frequencies in the space between the EBG holes and the waveguide. In order to avoid the unwanted resonances, wall corrugations in the waveguide are demonstrated to be an effective solution \cite{GlideSymm_holes1_2,Corrugations2}. Fig. \ref{figure14b} illustrates the corrugations implemented in the prototype to cancel field resonances.

The phase shifter and filter are formed by cascading ten holey unit cells. Using this number of unit cells, it is expected to obtain in the filter a rejection band from 61 GHz and in the phase shifter, a phase shift value around 180\textsuperscript{o} in the entire frequency range. The electric field distributions in the gap between layers, at different frequencies, for the reference, filter and phase shifter are illustrated in Fig. \ref{figure15}. Note that a higher field intensity along the waveguide width is observed at the middle of the three designs due to the narrowing of the waveguide height in this part. Moreover, it can be appreciated the absence of field leakage in the gap. At the lower frequencies (Figs. \ref{figure15a} and \ref{figure15b}), the input power reaches the output ports without visible attenuation. However, in both filter and phase shifter structures, the output signals are out-of-phase compared with the reference waveguide. At an intermediate operating frequency (Fig. \ref{figure15c}), the filter begins to produce effect since the selected frequency is in the stopband region (see Fig. \ref{figure9c}). On the other hand, the phase shifter still provides the same out-of-phase signal. Lastly, in Fig. \ref{figure15d}, the attenuation provided by the filter increases and the amplitude of the signal propagates in a shorter distance regarding the intermediate frequency. In contrast, at the output port of the phase shifter keeps arriving the 180\textsuperscript{o} phase shift, showing the low-dispersive behavior of the phase shifter.

\subsection{Experimental Validation}

Fig. \ref{figure16} illustrates the manufactured prototype, containing the reference waveguide, phase shifter and filter. The prototype was manufactured in CNC (Computer Numerical Control) technology. In order to measure the prototype, two setups have been employed since the R\&S-ZVA67 VNA is limited up to 66.5GHz. Therefore, ZVA-Z110E Converters in WR10 are used to reach up to 75 GHz. The produced insertion losses when the WR10 setup is applied are negligible. Fig. \ref{figure17} shows the simulated and measured results for the three manufactured designs. In simulations, aluminium is used instead of PEC in order to make a realistic comparison with the measurements. There exists a good agreement between the measurements and the simulation results, insertion loss being in all cases less than 2 dB. In the case of the filter, the rejection band reaches the 20 dB of attenuation from 63 GHz onwards. Note that the attenuation in the rejection band would be greater if a larger number of unit cells were employed to compose the filter. The peak in the attenuation is achieved around 70 GHz where the attenuation $\alpha$ in the mirror-symmetric unit cell has its maximum. This is in good agreement with the results presented in Fig. \ref{figure9c}, where the attenuation and phase constants of the same mirror-symmetric unit cell are analyzed.  Regarding the phase shifter, low insertion losses and good impedance matching band are obtained in a wide frequency range. The additional ripples in the reflection coefficient are produced by the coaxial to waveguide transitions used in the setup. In spite of that, good agreement between simulations and measurements in the $|S_{11}|$ are observed.

\begin{figure}[t!]
	\centering
	\subfigure[]{\includegraphics[width= 0.23\textwidth]{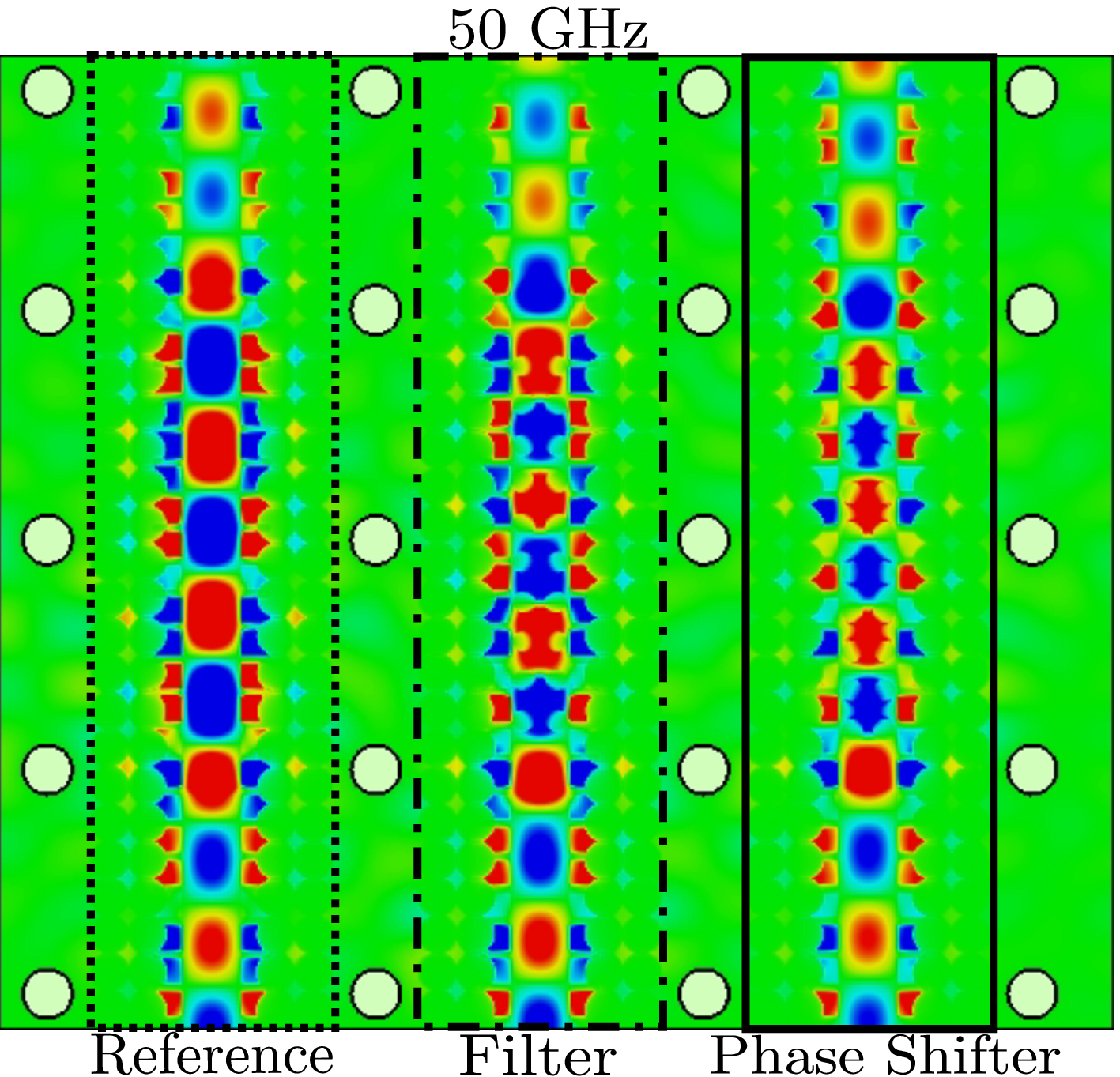}\label{figure15a}}
	\hspace{1mm}
	\subfigure[]{\includegraphics[width= 0.23\textwidth]{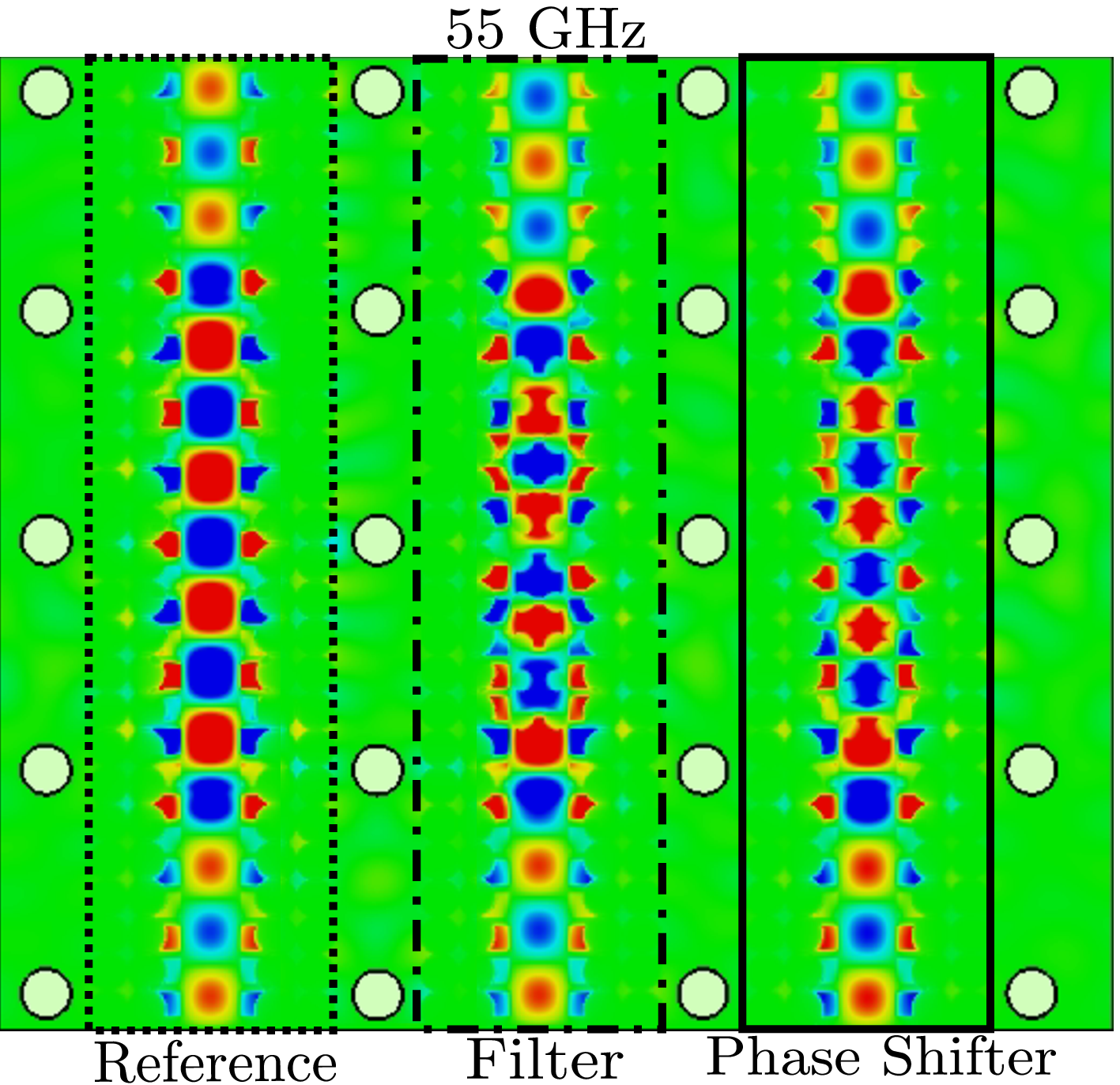}\label{figure15b}}
	\hspace{1mm}
	\subfigure[]{\includegraphics[width= 0.23\textwidth]{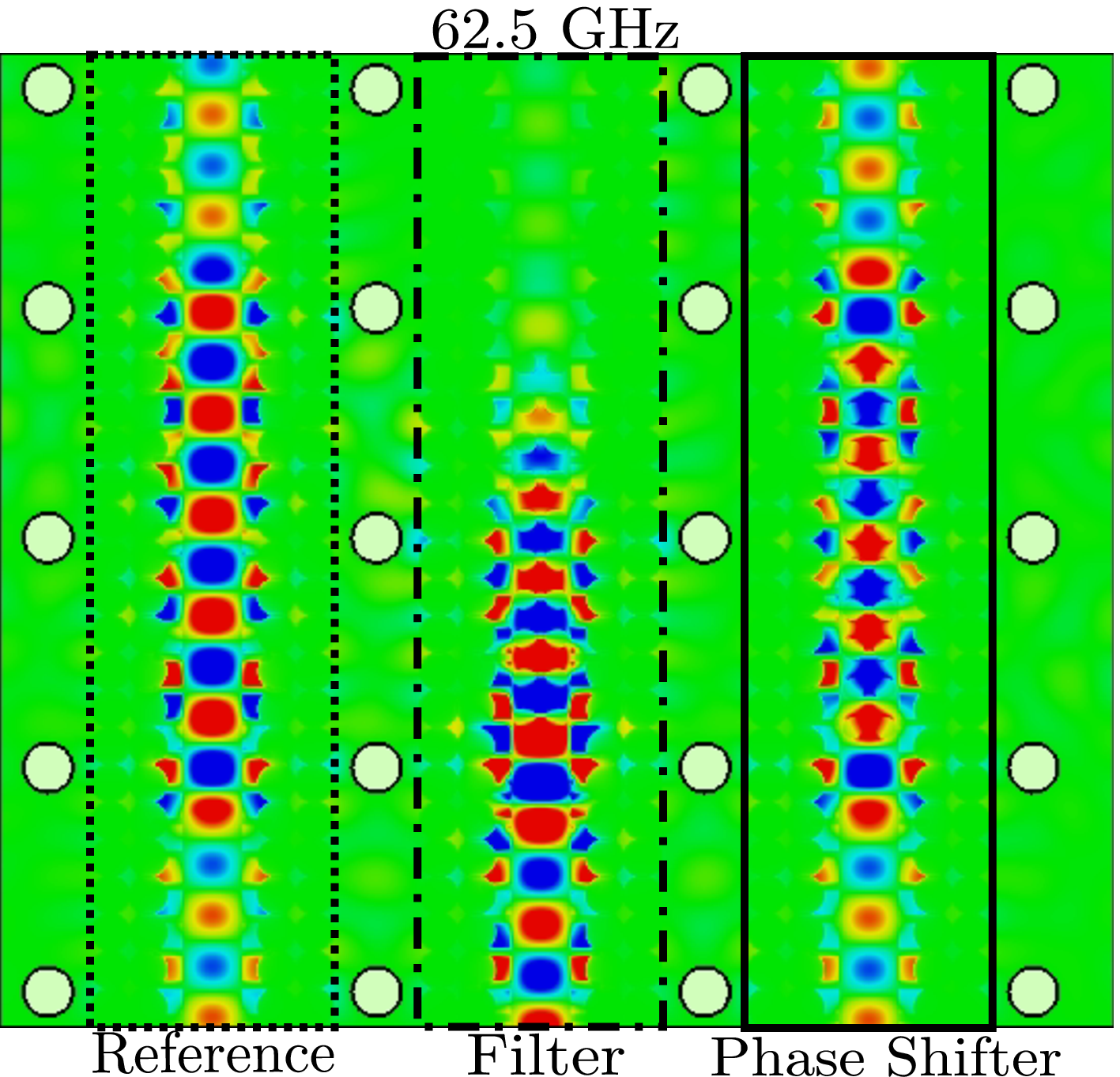}\label{figure15c}}
	\hspace{1mm}
	\subfigure[]{\includegraphics[width= 0.23\textwidth]{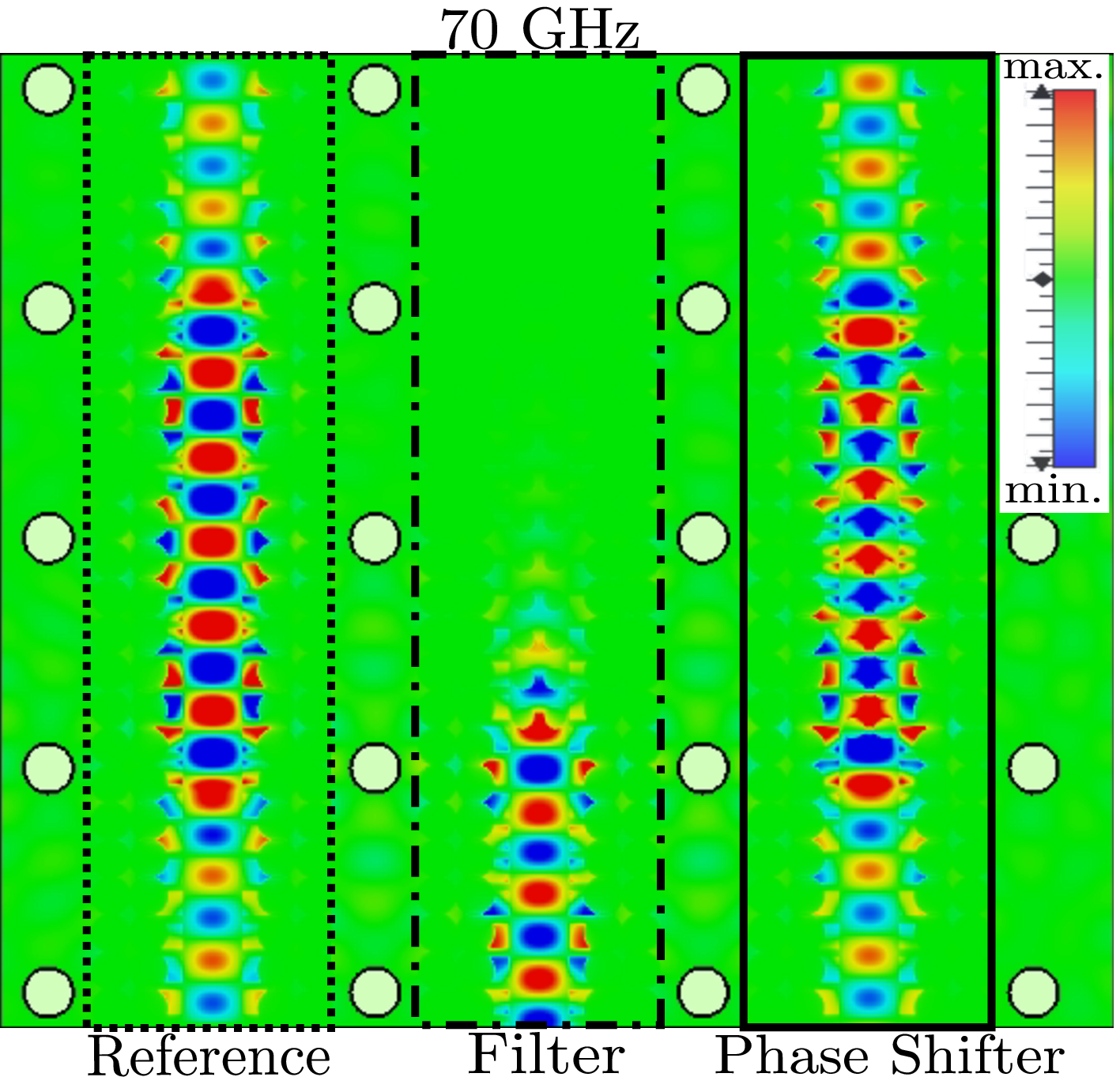}\label{figure15d}}
	
	\caption{Amplitude of the electric field distribution in the gap between layers at different frequencies in the operating band for the reference waveguide, filter and phase shifter at: (a) 50 GHz, (b) 55 GHz, (c) 62.5 GHz and (d) 70 GHz.}
	\label{figure15}
\end{figure}

\begin{figure}[t!]
	\centering
	\includegraphics[width= 0.48\textwidth]{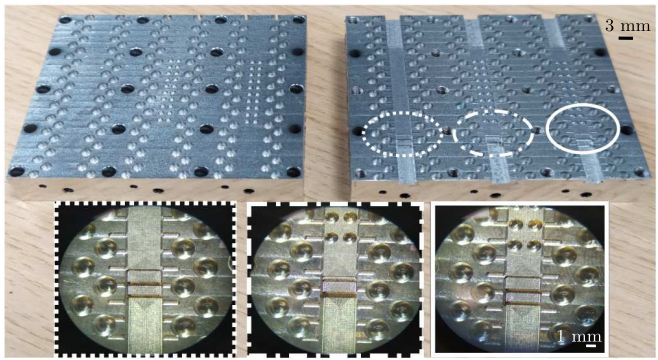}

	\caption{Forming layers of the manufactured prototype and details under microscope.} 
	\label{figure16}
\end{figure}

\begin{figure}[t]
	\centering
	\subfigure[]{\includegraphics[width= 0.49\textwidth]{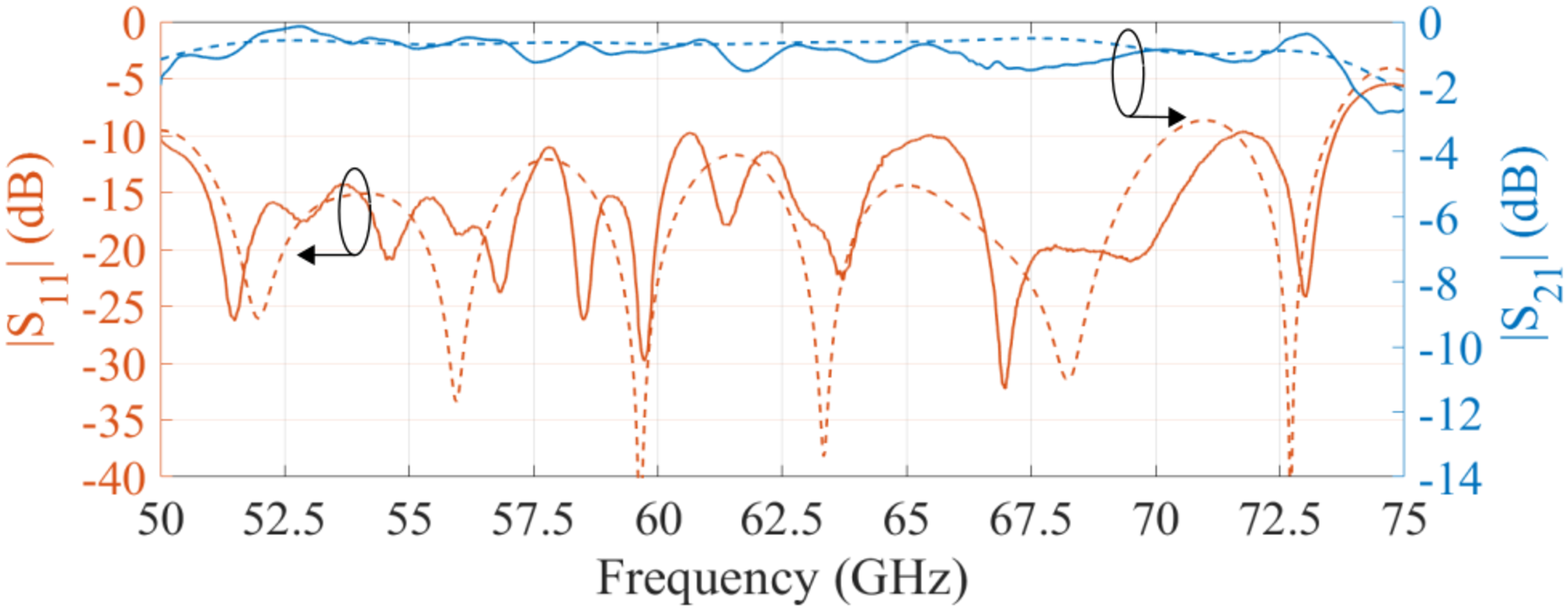}\label{figure17a}}
	\hspace{1mm}
	\subfigure[]{\includegraphics[width= 0.485\textwidth]{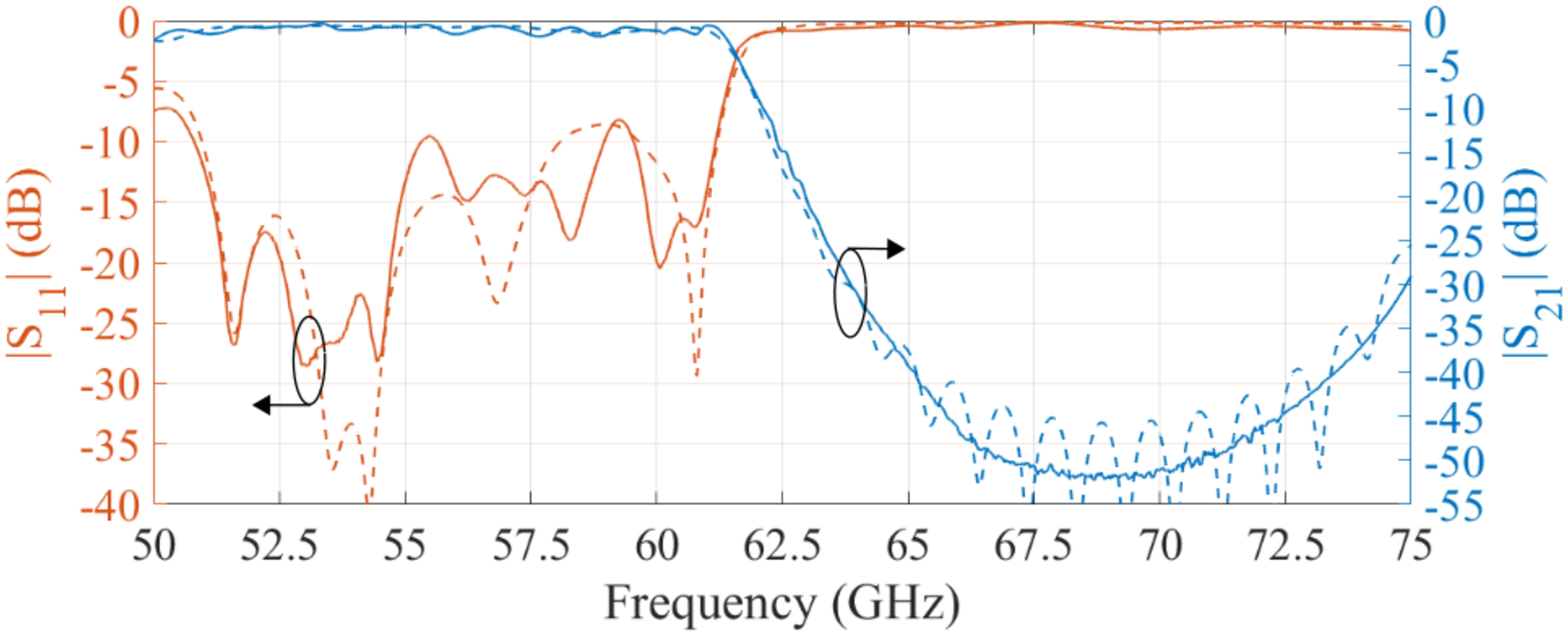}\label{figure17b}}
	\hspace{1mm}
	\subfigure[]{\includegraphics[width= 0.49\textwidth]{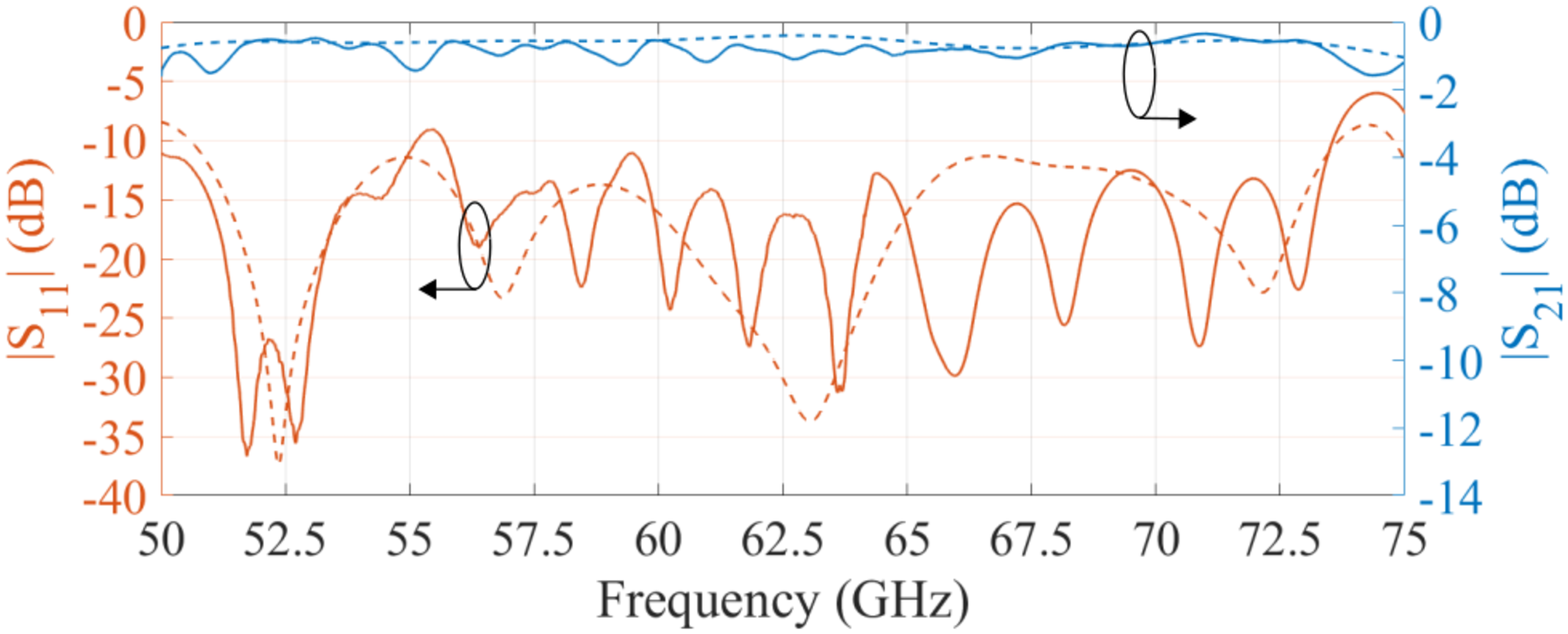}\label{figure17c}}
	
	\caption{Simulated (dashed lines) and measured (solid lines) results in magnitude of the S-parameters for the: (a) Reference waveguide, (b) Filter and (c) Phase shifter.} 
	\label{figure17}
\end{figure}

\begin{figure}[t]
	\centering
	\includegraphics[width= 0.45\textwidth]{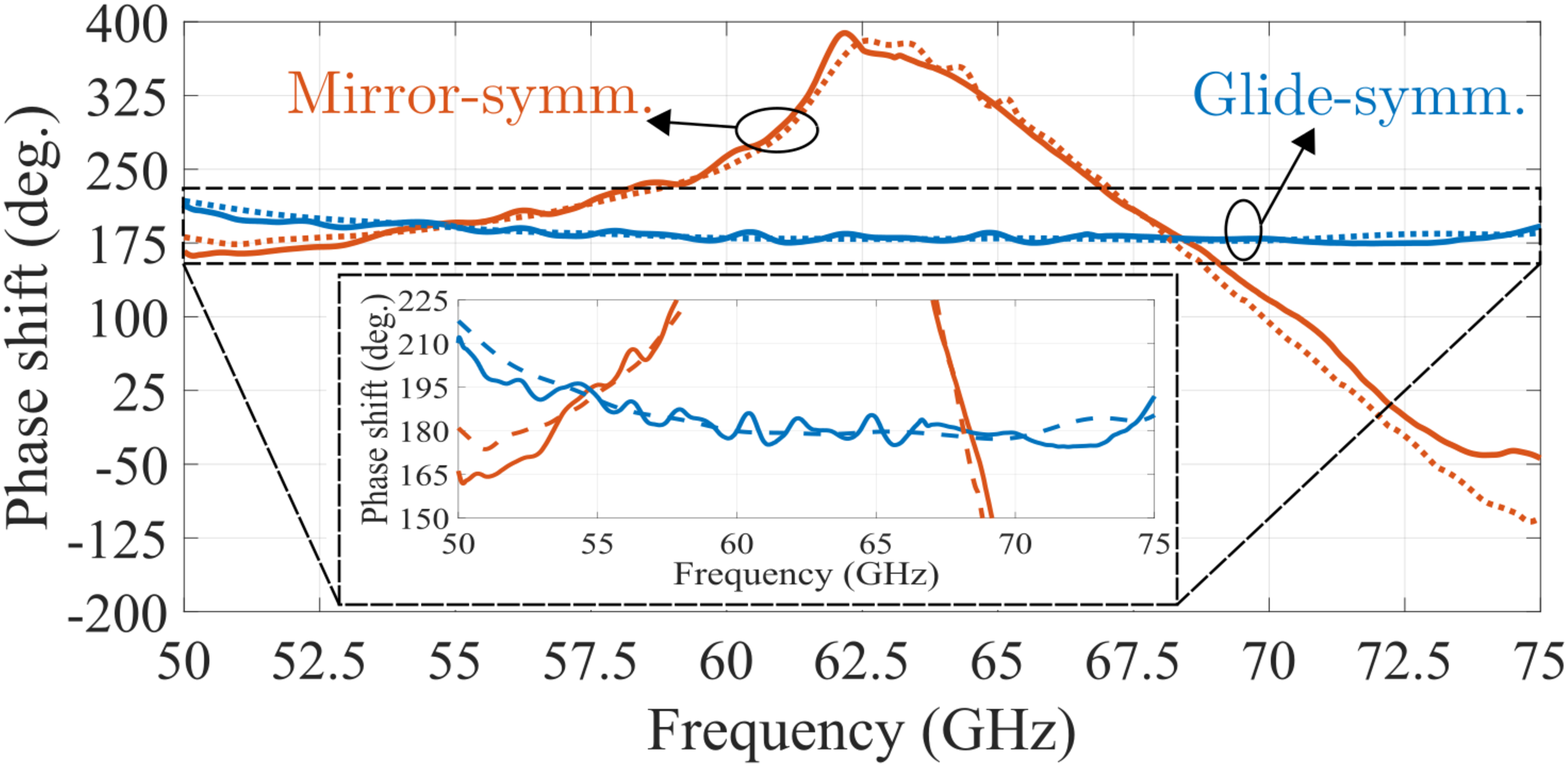}
	\caption{Simulated (dashed lines) and measured (solid lines) results in phase of the S-parameters for the phase shifter (in blue) and for the filter (in orange).}
	\label{figure18}
\end{figure}

Finally, the phase shift measurements for the phase shifter and filter are shown in Fig. \ref{figure18}. For the phase shifter, a 180\textsuperscript{o}$\pm$5\textsuperscript{o} is achieved from 56.5 GHz to 74.5 GHz (27.5\% frequency bandwidth). These results illustrate that a low-dispersive phase shifter (with respect to the reference waveguide) can be implemented with the use of holey waveguide whose holes are arranged in a glide-symmetric configuration. In addition, note that the dispersion of the filter (holey waveguide with mirror-symmetric holes) is higher out of the rejection band compared to the phase shifter (holey waveguide with glide-symmetric holes). This behavior is expected due to the approximated parallel fundamental mode of the holey waveguide in glide-symmetric configuration compared to the fundamental mode of the reference waveguide. At the lower part of the frequency band, there is an increase in the error of the produced phase shift due to two causes. First, at the lower frequencies of the stopband region of the EBG structure (50 GHz), the attenuation constant is low and thus, the electric field can propagate further in the transversal direction (see Fig. \ref{figure15a}). This fact produces an enlargement of the effective waveguide width that provokes a higher phase shift. When the frequency increases, the attenuation constant of the EBG structure is higher and the effective waveguide width becomes stable and similar to the one selected in the dispersion diagrams. Second, the first mode of the unit cell exhibits a lower dispersive behaviour beyond 54 GHz. Compared to the state-of-art hollow waveguide phase shifter \cite{Hedgehog_phaseShifter}, in this work it is obtained a wideband fully-metallic phase shifter with low-dispersive behavior and cost-effective design. Furthermore, it is of interest to compare the phase dispersion observed in the passband region of the filter with the phase shift produced by the phase shifter. As observed in Fig. \ref{figure18}, a lower dispersion is obtained when the holes are arranged in a glide-symmetric configuration (the case of the phase shifter) compared to a mirror-symmetric configuration (the case of the filter). This fact agrees with the results presented in Section II.A. In addition, good agreement is obtained in both the passband and rejection band of the filter, despite the fact that in the latter the amplitude signal received suffers from a great attenuation.

\section{Conclusion}

In this paper, the dispersion and filtering properties of the holey waveguides are analyzed in detail. Glide-symmetric holey waveguides  are demonstrated to be low dispersive in a large frequency bandwidth. On the other hand, both mirror-symmetric and broken glide-symmetric holey waveguides offer inherent stopbands that are beneficial for filter design. In addition, we show that the mirror-symmetric configuration provides a greater attenuation compared to the broken glide-symmetric one. A comparison between holey and pin-loaded waveguides is also carried out. The results show that periodic holey configurations are less frequency-dispersive compared to periodic structure with pins. The main properties of the holey waveguide with glide-symmetric and mirror-symmetric configurations are exploited to design a wideband phase shifter and a filter, respectively. They are prototyped in the WR15 waveguide by employing gap-waveguide technology. Moreover, the implementations of the phase shifter and filter provides an easy manufacturing since they are composed by holes that can be easily fabricated with drilling techniques in gap waveguides. The experimental results are in good agreement with the simulations. The phase shifter presents a phase shift of 180\textsuperscript{o}$\pm$5\textsuperscript{o} from 56.5 GHz to 74.5 GHz (27.5\% frequency bandwidth) and the filter offers a rejection band of 20-dB from 63 GHz to 75 GHz. The manufactured devices show cost-effective implementations of filters and low-dispersive phase shifters in millimeter-wave frequencies.


\begin{appendices}
\section{Multi-Modal Transfer-Matrix Method}
For the 1D periodic structure under study in this work, the eigenvalue problem that leads to the dispersion relation is presented in \cite{bloch_analysis1, multimodal_Paco} as
\begin{equation} \label{eigenproblem1}
    \mathbf{T} \begin{pmatrix} \mathbf{V} \\ \mathbf{I} \end{pmatrix} = e^{\gamma p} \begin{pmatrix} \mathbf{V} \\ \mathbf{I} \end{pmatrix}
\end{equation}
where 
\begin{equation}
\mathbf{T}= 
\bigs(
    \begin{array}{cc} \hspace*{-0.1cm}
        \strut{\overbrace{\strut\begin{matrix}
        A^{11} & \dots & A^{1N}\\
        \vdots & \ddots &  \vdots\\
        A^{N1} & \dots & A^{NN}
        \end{matrix}}^{\mathbf{A}}} 
        &
        \hspace*{-0.2cm}\bigstwo|
        
        \hspace*{0.3cm} \strut\smash{\overbrace{\begin{matrix} 
        B^{11} & \dots & B^{1N}\\
        \vdots & \ddots &  \vdots\\
        B^{N1} & \dots & B^{NN}
        \end{matrix}}^{\mathbf{B}}} \\
        
      \hline
        
        \hspace*{-0.12cm}  \strut\underbrace{\begin{matrix} 
        C^{11} & \dots & C^{1N}\\
        \vdots & \ddots &  \vdots\\
        C^{N1} & \dots & C^{NN}
        \end{matrix}}_{\mathbf{C}} 
        
        & 
        \hspace*{-0.18cm}\bigstwo|
        
        \hspace*{0.3cm} \strut\smash{\underbrace{\strut\begin{matrix} 
        D^{11} & \dots & D^{1N}\\
        \vdots & \ddots &  \vdots\\
        D^{N1} & \dots & D^{NN}
        \end{matrix}}_{\mathbf{D}}}
    \end{array}
\bigs)
\end{equation}
is the multi-modal transfer matrix of dimensions $2N \times 2N$, \textbf{V} and \textbf{I} are $N \times 1$ arrays containing the voltages and currents at the output ports, $\gamma=\alpha + j\beta$ is the propagation constant, $p$ is the period of the unit cell and $N$ is the number of considered modes.

If the structure under study is symmetrical and reciprocal, the $2N$-rank eigenvalue problem of \eqref{eigenproblem1} can be reduced to the $N$-rank eigenvalue problem \cite{multimodal_Paco}
\begin{equation} \label{eigenproblem2}
    \mathbf{A}\mathbf{V} =\cosh (\gamma p)\mathbf{V}
\end{equation}
which have two degenerate set of eigenvalues.

\end{appendices}

\ifCLASSOPTIONcaptionsoff
  \newpage
\fi


\end{document}